\documentclass[letterpaper,twocolumn,10pt]{article}
\usepackage{usenix2019_v3}

\usepackage[normalem]{ulem}
\usepackage[numbers]{natbib}

\usepackage[normalem]{ulem}
\usepackage{fancyvrb}
\usepackage{booktabs}
\usepackage{rotating}
\usepackage{graphicx}
\usepackage{textcomp}
\usepackage{xcolor}
\usepackage{multirow}
\usepackage{array}
\usepackage{dblfloatfix}
\usepackage{url}
\usepackage{placeins}
\usepackage{listings}
\usepackage{mdframed}
\usepackage{caption}
\usepackage{soul,color}
\usepackage{algorithm}
\usepackage{algpseudocode}
\usepackage[most]{tcolorbox}
\usepackage{xspace}
\usepackage{pifont}
\usepackage{mathtools}
\usepackage{colortbl}
\usepackage{enumerate}
\usepackage{makecell}
\usepackage{tikz}
\usepackage{subfig}

\newcommand{\memcached}{Memcached\xspace}
\newcommand{\redis}{Redis\xspace}
\newcommand{\memtier}{memtier\xspace}

 \newcommand{\bfs}{BFS\xspace}
 \newcommand{\pagerank}{PageRank\xspace}

 \newcommand{\xsbench}{XSBench\xspace}

\LetLtxMacro\origcite\cite	

\usepackage{pifont} 
\newcommand{\one}{\ding{182}\xspace}
\newcommand{\two}{\ding{183}\xspace}
\newcommand{\three}{\ding{184}\xspace}

\usepackage{mathtools}
\newcommand{\lat}{\texttt{Lat}}

\newcommand{\fault}{\texttt{Fault}}

\newcommand{\tco}{\texttt{TCO}\xspace}
\newcommand{\page}{\texttt{P}\xspace}
\newcommand{\comp}{\texttt{C}\xspace}
\newcommand{\hot}{\texttt{Hot\xspace}}
\newcommand{\usd}{\texttt{USD}\xspace}

\newcommand{\zswap}{{zswap}\xspace}

\newcommand{\skd}{TS-Daemon\xspace}

\definecolor{Gray}{gray}{0.85}
\newcolumntype{C}[1]{>{\centering\arraybackslash}p{#1}}
\newcolumntype{G}[1]{>{\centering\arraybackslash\columncolor{Gray}}p{#1}}
\newcolumntype{L}[1]{>{\raggedright\arraybackslash}p{#1}}
\newcolumntype{R}[1]{>{\raggedleft\arraybackslash}p{#1}}
\newcolumntype{D}[1]{>{\centering\arraybackslash} m{#1}}

\newcommand{\sepblock}{
\smallskip
\noindent
}

\usepackage{soul}

\newcommand{\twotier}{$2$-Tier\xspace}
\newcommand{\ntier}{TierScape\xspace}

\begin{document}

\title{Taming Server Memory TCO with \\Multiple Software-Defined Compressed Tiers}

\author{Sandeep Kumar, Aravinda Prasad, and Sreenivas Subramoney\\
Processor Architecture Research Lab, Intel Labs}

\date{}
\maketitle
\thispagestyle{empty}

\begin{abstract}
Memory accounts for 33--50\% of the total cost of ownership
(TCO) in modern data centers. 
We propose TierScape to tame memory TCO through the novel creation and judicious management of multiple software-defined compressed memory tiers.

As opposed to the state-of-the-art solutions that employ a 2-Tier solution, a single compressed tier along with DRAM, we define multiple compressed tiers implemented through a combination of different compression algorithms, memory allocators for compressed objects, and backing media to store compressed objects.
These compressed memory tiers represent distinct points in the 
access latency, data compressibility, and unit memory usage cost spectrum, allowing rich and flexible trade-offs between memory TCO savings and application performance impact.
A key advantage with \ntier is that it enables aggressive memory TCO saving opportunities by placing warm data in low latency compressed tiers with a reasonable performance impact while simultaneously placing cold data in the best memory TCO saving tiers. We believe TierScape represents an important server system configuration and optimization capability to achieve the best SLA-aware performance per dollar for applications hosted in production data center environments. 

TierScape presents a comprehensive and rigorous analytical cost model for performance and TCO trade-off based on continuous monitoring of the application’s data access profile. Guided by this model, TierScape takes informed actions to dynamically manage the placement and migration of application data across multiple software-defined compressed tiers. {On real-world benchmarks, TierScape increases memory TCO savings by 22\%--40\% percentage points while maintaining performance parity or improves performance by 2\%--10\% percentage points while maintaining memory TCO parity compared to state-of-the-art 2-Tier solutions}.

\end{abstract}
\begin{figure}
    \centering
    \includegraphics[width=1\linewidth]{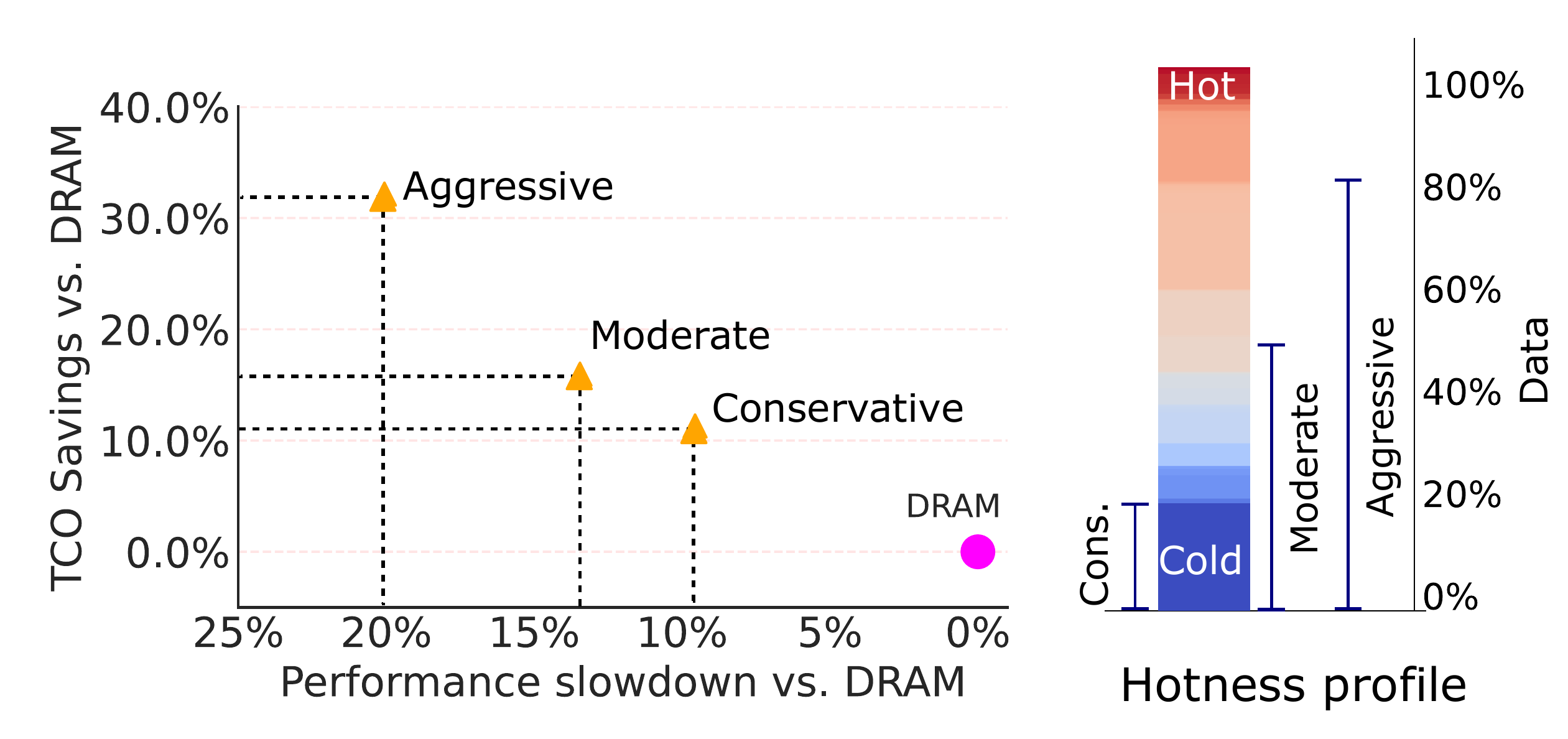}
    \caption{Memcached on a \twotier system (DRAM + a single compressed tier): {conservatively placing 20\% cold data in the compressed tier limits the memory TCO savings to 11\% with a 9.5\% slowdown. Placing around 50\% of data (including cold and some warm data) in the compressed tier results in 16\% memory TCO savings and 13.5\% slowdown. An aggressive approach that places around 80\% of data (including cold and most of the warm data) in the compressed tier results in 32\% memory TCO savings and 20\% slowdown.}
    }
    \label{fig:mod_teaser}
\end{figure}

\section{Introduction}
\label{sec:introduction}
Memory accounts for 33--50\% of the total cost of ownership (TCO) in modern data centers~\cite{tmo, src-study}. This
cost is expected to escalate further 
in order to serve the growing data demands of modern AI/ML applications whose working set already breaks the terabyte barrier~\cite{llm3,llm2,llm}, thus making it imperative to tame the data center's memory TCO.

The current state-of-the-art software-based solutions compress and place data in a compressed second-tier memory such as zswap in Linux to reduce memory TCO~\cite{gswap} (we refer to them as \twotier systems). Placing data in a compressed memory tier reduces the memory footprint of applications. As a result, systems can be provisioned with less memory, thus reducing the memory TCO in a data center.
However, memory TCO savings with compressed tiers is \textit{not free} as the data stored in such a tier must be decompressed before an application can access it, resulting in a performance penalty. 
Hence, to trade-off memory TCO savings and performance penalties, data center providers only place infrequently accessed or cold data in the compressed tier~\cite{gswap}.

We highlight the following critical observations and key limitations of the state-of-the-art \twotier systems. 
%
    \one On an average, 20--30\% of the data are cold in production systems~\cite{gswap, tmo, tpp, tmts, pond} and hence placing only cold data in second-tier compressed memory has limited memory TCO saving potential.
   \two Aggressively placing more data pages in a compressed second tier can increase memory TCO savings but results in a significantly higher and unacceptable performance penalty~\cite{tmts} (see Figure~\ref{fig:mod_teaser}). 
 \three Given the high cost of accessing data from a compressed tier, existing \twotier solutions do not compress warm pages, which accounts for 50--60\%~\cite{tmo,tpp} of the data pages, thus leaving significant memory TCO reduction opportunities on the table. 
%

In this paper, we seek to exploit memory TCO-saving opportunities beyond the cold data pages with an acceptable performance penalty. 
We propose \ntier, a novel solution with multiple software-defined compressed memory tiers (which we refer to as \textbf{N-Tier systems}) that dynamically manages placement and migration of data across compressed tiers to strike the best balance between memory TCO savings and application performance.
The compressed tiers can be a combination of 
different compression algorithms (e.g., lzo-rle, deflate, lz4), memory allocators for compressed objects (e.g., zsmalloc, zbud, z3fold), and backing media to store compressed objects (e.g., DRAM, non-volatile main memory~\cite{optane}, CXL-attached memory~\cite{cxl-website, cxl-samsung, cxl-micron}). 
TierScape's compressed tiers are distinct in access latency, unit memory usage cost, and capacity savings (compression ratio), enabling a holistic and flexible option space for hot/warm/cold data placement to balance memory TCO savings and application performance. TierScape thus compares very favorably to the rigid and restricted data placement and optimization space available in today's state-of-the-art 2-Tier systems.

TierScape, through its multiple compressed tiers, enables aggressive memory TCO saving opportunities by placing warm data pages in low-latency compressed tiers
with reasonable performance impact while simultaneously placing cold data in the best memory TCO saving tiers.
\ntier applies different placement and migration policies for warm/cold data based on the application's dynamic data access profile. For example, in our conservative model, which we refer to as the \textit{waterfall model} (\S\ref{sec:waterfall}), warm pages are initially placed in a low latency tier and eventually moved or aged to tiers with better TCO savings, thus progressively achieving better memory TCO savings.

TierScape introduces an advanced analytical model (\S\ref{sec:analytical_model}) that periodically recommends \textit{scattering} pages across multiple compressed tiers based on the access profile of the pages. The recommendations to move specific groups of pages to specific tiers are based on the usage patterns of the application's different memory regions, the relative costs of page access in different tiers, and the real-time memory TCO cost per tier incurred by the application. TierScape's multi-objective global optimization across application performance and memory TCO enables superior placement and control of hot/warm/cold page sets and calibrated maximization of performance-per-dollar metrics critical for data center operators.

The key contributions of the paper are as follows:
\begin{itemize}
\item To the best of our knowledge, we are the first to propose and demonstrate memory TCO savings for warm data with an acceptable performance impact.
\item Highlight the limitations with the state-of-the-art \twotier systems in saving memory TCO. Specifically, the limited TCO savings with cold data and its incapability to tap TCO saving opportunities for warm data with a reasonable performance penalty.
\item Demonstrate the benefits of defining multiple compressed memory tiers in the software that offer a rich and flexible trade-off between memory TCO savings and application performance impact. 
\item Judiciously manage page placement across tiers with waterfall and analytical models.


\end{itemize}

\section{Background}
\label{sec:background}


\subsection{Memory compression}
\label{sec:memory_tiers}
Linux kernel's zswap~\cite{zswap,gswap,zbud_zswap_kernel} supports memory compression where pages are compressed and placed in a compressed pool. Whenever a compressed page is accessed, \zswap decompresses the data from the compressed pool and places it in the main memory~\cite{zswap_kernel}. 
The Linux implementation of zswap has two key components: (i) the compression algorithm and (ii) the pool manager.

\sepblock
\textbf{Compression algorithms.}
The Linux kernel supports different compression algorithms such as deflate, lz4, lzo, and lzo-rle that differ in algorithmic complexity and the ratio of data compression achieved. However, \zswap is flexible enough to add new compression algorithms as required.
The deflate compression algorithm offers the best compression ratio but consumes comparatively higher CPU cycles to compress and decompress the data~\cite{lzo_vs_deflate, lzo,lz4}. On the other hand, lz4 is a fast compression algorithm but has relatively low data compressibility~\cite{lz4}. lzo (and its evolved variant lzo-rle) offers a balance between compression ratio and decompression overheads~\cite{lzo,lzo_rle_news,lzo_lwn_comment}. 
In addition, many compression algorithms such as lz4 have a ``level of effort'' parameter that can trade compression speed and compression ratio.

\sepblock
\textbf{Pool managers:} A pool manager manages how compressed pages are stored in \zswap. A pool is created in physical memory to store compressed data pages by allocating pages using the buddy allocator~\cite{buddy_allocator}. The pool dynamically expands to store more compressed objects by allocating more pages or contracts as required. To manage compressed objects inside the pool a custom memory allocator is used. Linux supports three pool memory allocators: zsmalloc, zbud, and z3fold~\cite{zbud_zswap_kernel,zsmalloc,z3fold}.

zsmalloc employs a complex memory management technique that densely packs compressed objects in the pool and thus has the best space efficiency. However, it has relatively high memory management overheads~\cite{zsmalloc}. 
zbud is a simple and fast pool management technique that stores a maximum of two compressed objects in a 4\,KB region. Due to this, the total space saved with zbud cannot be more than 50\%~\cite{zbud_zswap_kernel}. But, because of its simple object management, zbud has a relatively low memory management overhead.
z3fold is similar to zbud, but instead of two compressed objects, it can store three compressed objects in a 4\,KB region~\cite{z3fold}.

Linux allows users to pick a compression algorithm and a pool manager to manage \zswap. However, Linux supports only one active \zswap pool at a given time~\cite{zbud_zswap_kernel}. If a different compression algorithm or a pool manager is dynamically configured, the kernel creates a new pool and uses it to place compressed pages. The old pool is kept around till all data present in it is either faulted back to memory or invalidated~\cite{zbud_zswap_kernel}.     
\section{Motivation}
\label{sec:motivation}

\sepblock
\noindent \textbf{Missed opportunities for warm pages.}
Data center operators report that around 10--20\% of the data are hot and 20–-30\% of data are cold~\cite{gswap, tmo, tpp, tmts, pond}. This implies that around 50--70\% of the data pages are neither hot nor cold but can be considered as \textit{warm} pages. These warm pages can be (i) pages with relatively fewer accesses than hot pages or (ii) pages that are transitioning from hot to cold as hot data does not become cold instantaneously but rather follows a gradual process where it ages itself to cold. However, a cold or warm page can instantaneously become hot, depending on the access pattern of the application.
Existing \twotier solutions do not consider exploiting warm pages for compression, thus missing significant memory TCO-saving opportunities. 


\sepblock
\noindent \textbf{Drawbacks with aggressive data placement.}
A naive approach to aggressively place more data in the compressed second-tier memory to increase memory TCO savings results in a significantly higher and unacceptable performance penalty (Figure~\ref{fig:mod_teaser}).
%
However, replacing a highly compressible tier with a low compression, low access latency tier (due to low decompression latency) can enable aggressive data placement in the compressed tier. However, it severely impacts the memory TCO savings due to low compression ratio. 

Employing page prefetching~\cite{gswap} that prefetches or decompresses pages from compressed memory can mitigate high-performance penalty to the extent of prefetching accuracy. However, pages that the prefetcher fails to identify for prefetching still incur high access latency when accessed, and incorrectly prefetched pages results in decreased memory TCO savings. 
Nevertheless, prefetching can be additionally employed in an N-Tier memory context and we note it as a future work of interest for the systems community.

\begin{figure}
    \centering
    \includegraphics[width=1\linewidth]{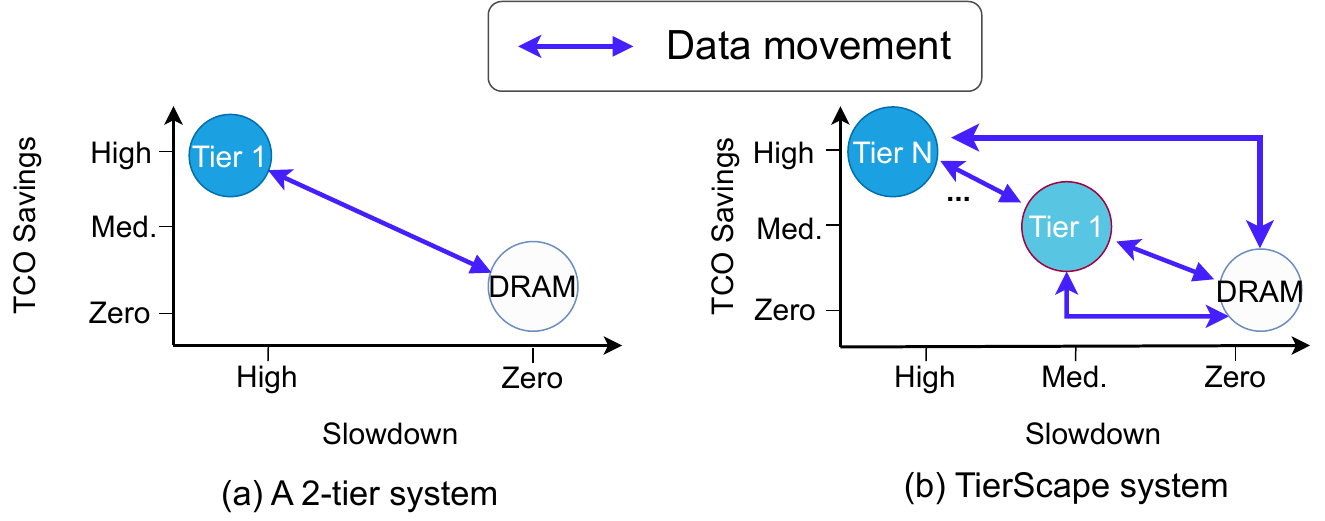}
    \caption{Data placement options in \twotier and N-Tier systems} 
    \label{fig:two_tier_system}
\end{figure}

\sepblock
\noindent \textbf{Limited placement choices.}
To reiterate the central observation here: the key limitation with \twotier systems is the binary decision options they face for data placement --  either in DRAM or in the compressed second tier (Figure~\ref{fig:two_tier_system}).
This severely limits the flexibility and choices for page placement towards a better balance between application performance and memory TCO.

\sepblock
\noindent \textbf{Summary.}
To conclude, the current 2-Tier approaches fail to exploit the temperature gradient that naturally manifests across a large population of the application's pages over time to simultaneously achieve better TCO and performance.

\section{Concept}
\label{sec:concept}

The core concept behind our proposal is to define multiple compressed tiers in the software. Each compressed tier is defined through a combination of (i) compression algorithms, (ii)  memory allocator for the compressed pool, and (iii) different backing media -- each providing a different access latency and memory cost per byte, as we discuss below.

\sepblock
\noindent \textbf{Compression algorithms.}
Compression algorithms with low compression ratio and, consequently, a low decompression latency are suitable for low latency tiers, but they provide only marginal memory TCO savings. Whereas other compression algorithms, such as deflate with high compression ratio and, consequently, high decompression latency, are suitable for high memory TCO savings tiers but with significantly high memory access latency.

\sepblock
\noindent \textbf{Pool allocators.}
As zsmalloc densely packs compressed objects in the pool, it is suitable for high memory TCO saving tiers, but it has high memory management overheads, thus impacting the decompression latency. zbud, with its simple and fast pool management, is suitable for low latency tiers but is less space efficient, resulting in tiers with low memory TCO savings.

\sepblock
\noindent \textbf{Physical media.}
The access latency of the storage medium where the compressed pages are stored is crucial for the performance of the tier. Storing compressed pages on DRAM offers the lowest possible media access latency~\cite{yang2020empirical} and hence suits low latency tiers. But doing so also reduces the overall memory TCO savings potential. Using cheaper and denser memory, such as NVMMs or CXL-attached memory, to store compressed pages increases memory TCO savings but adds to decompression latency, rendering them attractive for use as high memory TCO saving tiers.

The key idea for enabling aggressive memory TCO savings is to use tiers with low latency for warm pages that can save memory TCO at moderate performance overheads. Meanwhile, tiers with high compression ratios and high access latency are used for cold pages.

\begin{table}
    \centering
    \footnotesize
    \caption{Different options available in Linux for setting up a compressed tier}
    \label{tab:zswap_config}
    {%
    \begin{tabular}{|C{.35\linewidth}|C{.19\linewidth}|C{.3\linewidth}|}
    \hline
    \textbf{Compression algorithm}  & \textbf{Allocators}            & \textbf{Backing media}                         \\ \hline
     Deflate, LZO, LZO-RLE, LZ4, Zstd, 842, LZ4HC & zsmalloc, zbud, z3fold & DRAM, CXL-attached memory, NVMM \\ \hline
    \end{tabular}%
    }
\end{table}

\subsection{Characterization of compressed tiers}
\label{sec:characterization_zswap}

We start by comparing the access latencies and memory TCO benefits of compressed tiers with different configurations in Linux. The Linux kernel offers only two configuration parameters for a zswap compressed tier (compression algorithm and pool manager) but does not offer any control over where the pool is allocated, i.e., the kernel cannot be instructed to allocate the pool on DRAM or NVMM. 
We modify zswap to add a configuration parameter -- \textit{backing media}, that specifies from which hardware media the pages for a particular compressed pool are to be allocated. \textit{This allows us to construct tiers specifying backing media. }

The latency of decompressing a page from \zswap is primarily dominated by the compression algorithm, pool manager, and backing media. 
With the available choices in Linux (as shown in Table~\ref{tab:zswap_config}), we can create a total of 63 different \zswap compressed tiers ($C^7_1 * C^3_1 * C^3_1$).
In addition, the compressibility ratio and decompression latency of a given tier also depend on the compressibility of the input data.

In order to allow for multiple operating points in the space of access latency and memory TCO savings, we define 12 tiers configured based on widely used compression algorithms and pool managers. We initialize 10\,GB of data in memory, compress and place them in a compressed memory tier and then access them. We repeat this experiment for all 12 tiers.
To characterize with different input data, we use two data sets from the Silesia corpus~\cite{silesia}, \textit{nci} and \textit{dickens}, with the former being more compressible~\cite{silesia_compression}. 
We measure the access latency and compression ratio.

  \begin{figure}[!ht]
    \centering
     \subfloat[Access latency\label{fig:nci_mt_load_mt}]{
        \includegraphics[width=0.8\linewidth]{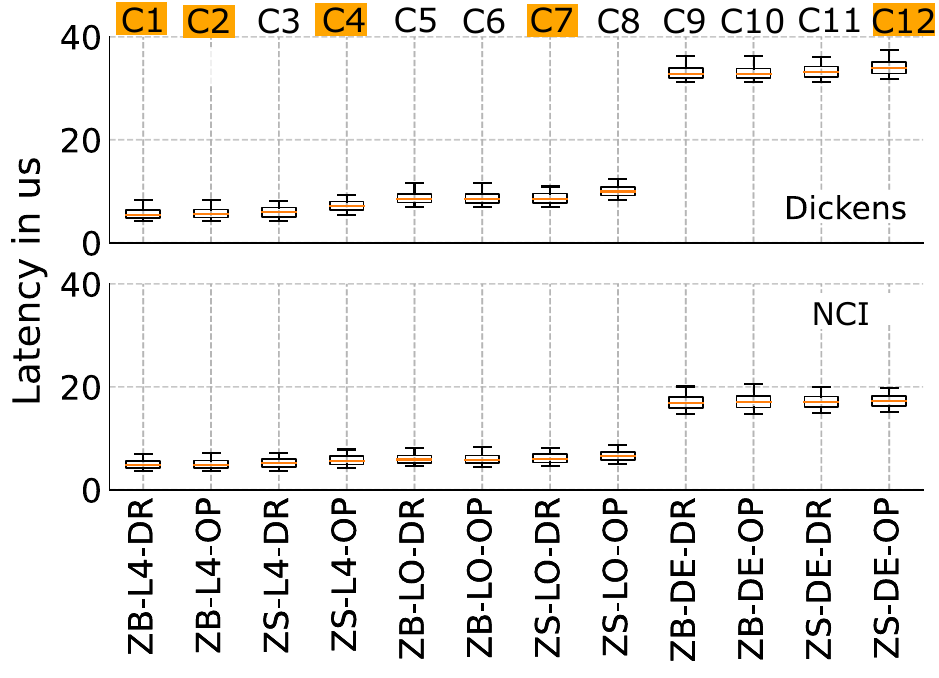}
    }

     \subfloat[Memory TCO savings\label{fig:tco_cost}]{
    \includegraphics[width=0.8\linewidth]{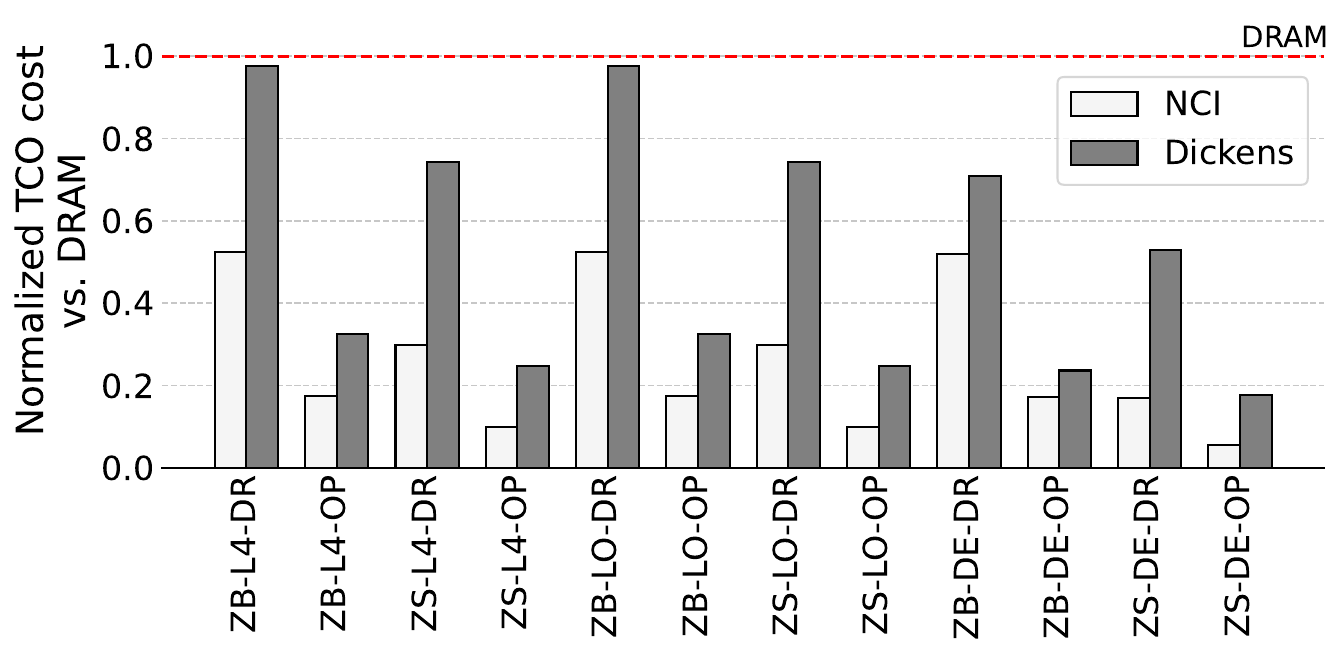}
    }
    \caption{Characterization results for 12 different software-defined compressed tiers for \textit{dicken} and \textit{nci} data sets. Encoding: \textbf{ZS, ZB} refers to zsmalloc and zbud pool managers, respectively. \textbf{L4, LO, DE} refers to lz4, lzo, and deflate compression algorithms, respectively. \textbf{DR, OP:} refers to DRAM and Optane~\cite{optane} as the backing storage media, respectively.}
\end{figure}

\subsubsection{Access latency}
Figure~\ref{fig:nci_mt_load_mt} shows the access latency for both \textit{nci} and \textit{dickens} data sets. Access latency with the lz4 algorithm is the fastest, followed by lzo, and lastly, deflate. 
As expected, the performance of zbud pool manager is also better than zsmalloc. This is because zbud employs a simple algorithm that enables faster page lookup.
Finally, the access latency of DRAM-backed tiers is better than those backed by the Optane~\cite{optane} due to the higher media access latency in the latter~\cite{optane_characterization}.

\subsubsection{Memory TCO savings}
Figure~\ref{fig:tco_cost} shows the normalized memory TCO savings of compressed tiers relative to uncompressed data in DRAM. Total TCO savings depend on data compressibility, compression algorithm, and backing media. 
The cost per gigabyte for storing data on Optane is typically 1/3 $\sim$ 1/2 of the cost of storing data on DRAM~\cite{flexhm}. 
Hence, the memory TCO for Optane-backed tiers is lower than that of DRAM-backed tiers.

Furthermore, for tiers using the same compression algorithm and backing media the TCO savings depend on the memory allocator for the compressed pool manager. For example, a tier using zsmalloc as its pool manager has a lower memory TCO than a tier using zbud. This is because zsmalloc can pack compressed objects more tightly. 
Finally, the deflate compression algorithm offers the best compression ratio.

 
\subsection{Tiers selection methodology}
\label{sec:tiers_selection}
In order to illustrate the flexibility and robustness of the TierScape proposal, we select five compressed tiers (C1, C2, C4, C7, and C12) that we define in the software. 

We pick C1 and C12 as they offer the best performance configuration and best memory TCO savings configuration, respectively. Other tiers with deflate compression algorithms offer a similar performance latency without additional TCO benefits, and hence we do not select any other deflate-based tiers.
We select C2 as it offers the lowest latency for an Optane-backed compressed tier. C1 and C2 use zbud and lz4 as their pool manager and compression algorithm -- restricting the compression ratio to 2. Hence, we select C4, which uses a fast compression method (lz4), tightly packs compressed objects (due to zsmalloc), and is stored on low-cost Optane. Finally, we select C7, which fills the gap between access latency and memory TCO savings. 
We use this set of tiers for our experiments to demonstrate rich and flexible placement opportunities. 
\newcommand{\userinp}{$\alpha$\xspace}
\newcommand{\mts}{MTS\xspace}

\section{Data placement in \ntier}
\label{label:polices}

In this section, we present two distinct data placement models that fully exploit the benefits of N-Tier systems. 
Note that we develop these models to show rich and flexible data placement options to tame memory TCO.
However, we believe that having multiple software-defined compressed tiers opens up a plethora of exploration opportunities for innovative data placement policies.

For ease of our discussion, we assume the system is configured with DRAM + $N$ compressed tiers. Furthermore, the tiers are ordered from low latency to high latency (and, consequently, low TCO savings to high TCO savings), i.e., Tier 1 offers the best performance but with the least memory TCO savings. In contrast, Tier $N$ offers the best memory TCO savings with high performance impact.

\subsection{Waterfall model}
\label{sec:waterfall}
\begin{figure}
    \centering
    \includegraphics[width=.6\linewidth]{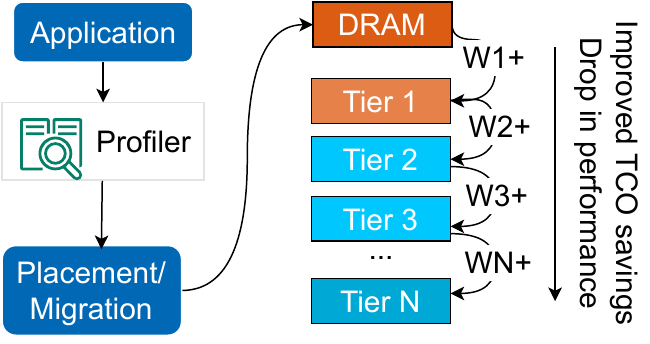}    
    \caption{Page placement with the N-Tier waterfall model}
    \label{fig:waterfall_model}
\end{figure}
A \twotier memory TCO saving solution uses a hotness threshold ($H_{th}$) to decide which pages should be pushed from DRAM to the compressed tier. As seen before, an aggressive threshold (a high value for $H_{th}$) pushes more pages to the compressed tier saving  additional memory TCO but at the cost of high performance penalty due to multiple high latency page faults from the compressed tier. The waterfall model extends this approach naturally to leverage multiple software-defined tiers available to achieve better memory TCO savings while limiting the performance penalty.

The model starts by monitoring the pages accessed by an application for a fixed duration -- henceforth referred to as a \textit{profile window}. 
As shown in Figure~\ref{fig:waterfall_model}, at the end of each profile window, all the pages that have a \textit{hotness value} (access count) less than the threshold ($H_{th}$) are moved from DRAM to low-latency tier T1. 
This reduces the total memory TCO upfront as some data pages have been placed in a compressed tier (albeit the tier has a low compression ratio). The advantage is that these compressed pages can be decompressed and placed in DRAM when accessed without high performance penalty as T1, by design, is a low latency compressed tier.

During the next profile window, some pages will be faulted back to DRAM from T1 as per the application's access pattern. Once the profile window ends, all the pages that are still in T1 are, in fact, getting colder as they were not accessed in the last profile window. The model moves (or waterfalls) all the data from T1 to T2. This further increases the memory TCO savings as T2 is better than T1 in memory TCO savings. 

At the end of each profile window, the model waterfalls all the data in all the tiers to one tier below it (to a higher TCO saving tier), except for the last tier. However, pages that are accessed by the application are decompressed and placed in DRAM, irrespective of its tier and these pages have to start the journey again from T1.

\sepblock
\noindent \ul{Benefits:}

\sepblock
    \textbf{Upfront memory TCO savings.} The memory TCO savings start upfront, as all the cold and warm data can be immediately placed in low latency compressed tiers without significant performance impact.
    
\sepblock \textbf{Tolerate profiling inaccuracies.} Existing profiling techniques such as PMU's~\cite{pebs} do not provide a 100\% accurate memory access profile~\cite{hemem} which can result in incorrectly identifying a hot or a warm page as a cold page. The penalty for placing a hot page incorrectly classified as a cold page in a \twotier solution can be significant~\cite{tmts}. As the waterfall model initially places all pages, including incorrectly classified hot or warm pages, in low latency compressed tiers, it incurs a minimal performance penalty when they are accessed.
        
    \sepblock \textbf{Gradual convergence to maximum TCO savings.} Waterfall model gradually moves cold pages to better memory TCO saving tiers with each profile window. Hence, it eventually converges to a stable phase where all the cold data pages are placed in the best memory TCO saving tier, thus maximizing the TCO savings. 
    

\sepblock
\ul{Limitations:}

    \sepblock  \textbf{Cold page convergence. } Cold pages (i.e., pages with 0 access count) requires $N$ profile windows (in an $N$-Tier setup) to converge to the last or best TCO saving tier. It misses the opportunity to aggressively place cold pages directly in best memory TCO saving tiers. 
    
    \sepblock \textbf{Limited flexibility to fine-tune page placement.} Waterfall model does not offer flexibility to fine-tune page placement.  A hotness threshold parameter fully controls page placement. For example, users cannot specify placement criteria or requirements to trade off memory TCO savings and performance penalties.
   


\subsection{TierScape's analytical model}
\label{sec:analytical_model}

\begin{figure}
    \centering
    \includegraphics[width=.7\linewidth]{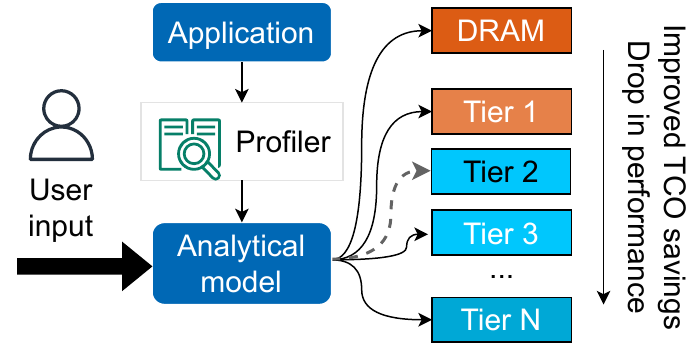}    
    \caption{Page placement with the N-Tier analytical model. }
    \label{fig:ilp_model}
\end{figure}

We propose an analytical data placement model
to address the limitations of the waterfall model. Analytical model can directly distribute data ( Figure~\ref{fig:ilp_model}) to different memory tiers based on the hotness profile of the data. In addition, the model provides fine control to the users to balance the trade-off between memory TCO savings and performance penalty by exposing a user-guided tunable \textit{``knob''}.

\begin{figure}
    \centering
    \includegraphics[width=.9\linewidth]{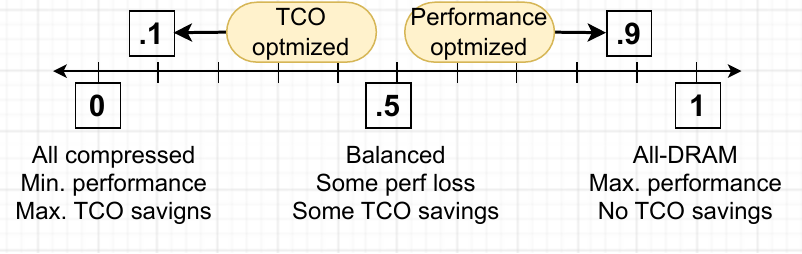}
    \caption{The memory TCO is minimal when all the data is placed in a highly compressible tier, while it is maximum when all the data is in DRAM. The difference between the two is the TCO saving opportunity that is  tuned with a knob in the analytical model.}
    \label{fig:tco_perf_scale}
\end{figure}

As shown in Figure~\ref{fig:tco_perf_scale}, the range of the knob is [$0$, $1$]. A value of $1$ indicates the model is tuned for maximum performance, which results in zero memory TCO savings as all data pages are placed in DRAM. On the other hand, a value towards $0$ indicates that the model is tuned to maximize TCO savings while striving to minimize performance penalty. 

\subsubsection{Data placement modeling}
The analytical model is initiated with a knob value -- say \userinp $\in [0,1]$.
The theoretical memory TCO savings achievable is the difference between TCO$_{max}$ -- when all the data is in DRAM and TCO$_{min}$ -- when all the data is in the last tier. The maximum TCO savings (or \mts) can be defined as follows:
 \begin{equation}
\text{MTS}  = \text{TCO}_{max}- \text{TCO}_{min}
 \end{equation}
The analytical model can be tuned to achieve TCO savings within [0, \mts] by configuring \userinp.

At the end of each profile window, the model uses \userinp and the profiled data to  solve the following:
\begin{equation}
    \begin{array}{ll@{}ll}
    \text{minimize}  & perf\_{ovh}_{NT} &\\
    \text{subject to}&  \tco \le (\tco_{min} + \alpha*\text{MTS})  &
    \end{array}
    \label{eq:main_eq}
\end{equation}

In order to solve Equation~\ref{eq:main_eq}, we start by formally defining performance overhead ($perf\_{ovhd}$) and the memory TCO.

\subsubsection{Modeling performance overheads}
In terms of memory accesses, an application executes optimally when all its load operations are directly from DRAM (instead of a compressed tier). Let us refer to this performance as $perf_{opt}$.

Consider a scenario when a few of the application's pages are placed in a single compressed tier $T_i$.
If an application attempts to read those pages, it will result in $\fault_{T_i}$ faults, with each fault incurring 
$\lat_{T_i}$ latency overheads to decompress the data.
Once the pages are decompressed, the accesses are served from DRAM. Hence the performance with a compressed memory tier includes the cost of accessing memory regions from DRAM:
\begin{align}
    perf''  &= perf_{opt}+\fault_{Ti}*\lat_{Ti} \\
    perf\_{ovh} &= perf^{''} - perf_{opt} \\
    &=\fault_{Ti}*\lat_{Ti} 
\end{align}
Here, $perf\_{ovh}$ is the performance overhead due to accessing pages in a compressed memory tier $T_i$ and is equal to the total time spent serving the faults from Tier $T_i$. 

Generalizing this when $N$ compressed tiers are used, the performance overhead ($perf\_{ovh}_{NT}$) can be defined as:
\begin{align}
    perf\_{ovh}_{NT} &= \sum_{y=1}^{N}(\fault_{Ty}*\lat_{Ty})
\label{eq:tco_eq}
\end{align}
%
Here, $\fault_{Ty}$ is the number of faults the application incurs from a compressed tier $T_y$. 
However, the model does not have this information while making the placement decision at the end of the current profile window, as it cannot estimate the number of future faults for the application.

In order to estimate the number of faults, the model exploits the fact that for an application with stable access patterns,
the total number of faults to a data region ($r$) in the next profiling window, if placed in a compressed tier, will be proportional to the hotness of the data region ($\hot_r$) in the previous profiling window. Hence, $\fault_{Ty}$ is proportional to the sum of the hotness of all the data regions stored in that tier $T_y$:
\begin{equation}    
\label{eq:fault_hotness_map}
    \fault_{Ty} \propto  \sum_{r=1}^{R}\hot_{r}
\end{equation}

Hence, we use the following equation, which is in terms of page hotness from the previous profile window, to estimate the performance overhead:
\begin{align}
\label{eq:final_perf_ovh}
\Aboxed{
    perf_{ovh} &= \sum_{y=1}^{N}\left(\left(k_{y}\sum_{r=1}^{R}\hot_{r}\right)*\lat_{Ty}\right) 
    }
\end{align}
Here, $k_y$ is a constant factor. For the rest of the paper, we use $k_y$ as 1.


\subsubsection{Modeling memory TCO}
The memory cost of placing data on a particular tier depends on the backing media and compressibility of data. 
The memory TCO is highest when all the data (measured as 4\,KB pages, $\page_{tot}$) of an application is in DRAM and is defined as:
\begin{align}
    \tco_{max} &= \page_{tot}*\usd_{DRAM}
\end{align}
Where $\usd_{DRAM}$ is the cost of storing a single 4K page in DRAM.
Similarly, the memory TCO is lowest when all the data is placed in the best memory TCO savings tier ($N$):
\begin{align}
    \tco_{min} &= \page_{tot}*(1/C_{T_N})*\usd_{T_N} 
\end{align}
Where $\usd_{T_N}$ is the cost of the media backing the compressed tier $T_N$.
$C_{T_N}$ is the compressibility ratio of tier $T_N$ defined as:
\begin{equation}
C_{T_N} = \frac{\text{Original size of data on $T_N$}}{\text{Compressed size of data on $T_N$}}    
\end{equation}
As discussed before, the compressibility of a tier depends on the compression algorithm, pool manager, and the data.

In an N-Tier system, the memory TCO can be defined as the sum of the cost to store data in DRAM and the cost to store data in compressed tiers. It can be defined as:
\begin{align}
\label{eq:tcocost}
\boxed{
    \tco_{NT}= \page_{DRAM}*\usd_{DRAM}+\sum_{y=1}^{N}(\page_{Ty}*(1/\comp_{Ty})*\usd_{Ty})
}
\end{align}
Here, $\page_{Ty}$ is the number of pages placed in Tier $T_y$.
We use Equation~\ref{eq:final_perf_ovh} and Equation~\ref{eq:tcocost} to solve Equation~\ref{eq:main_eq} as an \textit{integer linear program} (or ILP). The number of pages in DRAM $P_{DRAM}$ and pages in each compressed tiers $P_{T_y}$ 
are the optimization variables. The model outputs the final placement of pages that satisfy the constraints. We then place data on different compressed tiers as per the model's recommendation.


\subsubsection{Discussion}
The model quickly converges to optimal data placement based on the profiled hotness of the data. Cold data are directly placed in the most optimal tier as per the constraints instead of "waterfalling" on multiple tiers.
In addition, the user-guided tunable knob ($\alpha$) 
enables fine-tuning memory TCO and performance penalty trade-off.





\section{Implementation}
\label{sec:implementation}


\begin{figure}
    \centering
    \includegraphics[width=.7\linewidth]{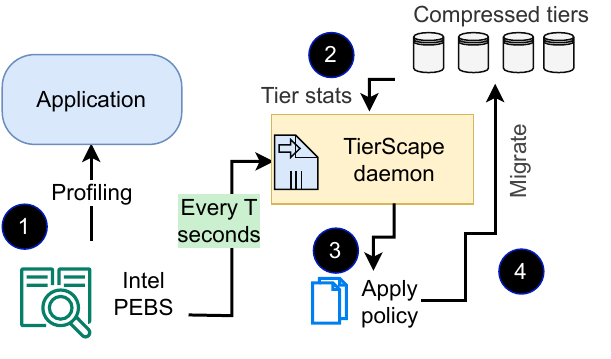}
    \caption{A high-level working of \ntier} 
    \label{fig:ntier_high_level}
\end{figure}

\subsection{Linux kernel changes}
\label{sec:impl_kernel_changes}

\textbf{Tier's backing media:}
As discussed in Section~\ref{sec:characterization_zswap}, the Linux kernel configures a compressed memory tier using two parameters: the compression algorithm and the pool manager~\cite{zswap}. 
We augment the \zswap subsystem to add a third parameter to specify a \textit{backing media} which can be NVMM or CXL-attached memory. 
We enhance the kernel to allocate physical memory only from these backing media when the pool is created or when the pool dynamically expands to store more compressed objects.
 
\sepblock
\textbf{Multiple active compressed tiers:}
Linux kernel only supports a single active \zswap tier. Upon creation of a new compressed tier, all new data compression requests are directed to the newly created tier. The kernel deletes the old tiers if they are empty. We modify the \zswap subsystem to support multiple active compressed \zswap tiers and also allow multiple compressed tiers to co-exist.  

\sepblock
\textbf{API changes:}
Once we setup multiple active compressed tiers, the challenge is to instruct the kernel to send a specific set of pages to a target tier based on the model recommendation. For example, the model can recommend placing a few pages in Tier T2 and a few others in T4. To ensure the placement of pages in the recommended target tier, we augment the \texttt{struct page} structure with a $tier\_id$ field which is updated by a modified \texttt{madvise()} function. During page compression, the \zswap module reads this field and places the compressed page in the intended tier. 

The decompression operation remains unchanged. During a page fault, the handle in the page table entry is used for the RB-Tree lookup to find the associated swap entry. The swap entry contains the tier information, including the pool details and other relevant information to handle the fault~\cite{zswap}.

\sepblock
\textbf{Page migration between tiers:}
We enhance the kernel to allow the migration of pages between two compressed tiers.
Currently, we follow a naive approach while migrating pages between compressed tiers by first decompressing the page from the source tier and then compressing again and placing it in the destination tier. 
This can be further optimized by skipping the decompression step if the source and destination tiers use the same compression algorithm.

\sepblock
\textbf{Tiers statistics:}
We added support in the \zswap subsystem to collect per-tier statistics such as the number of pages in the tier, size of the compressed tier, and total faults. 

\subsection{\skd}
\label{sec:skd}
As shown in Figure~\ref{fig:ntier_high_level}, we implement our \ntier logic as a daemon (\skd). \skd uses the hardware counters to profile the memory access pattern of an application for a fixed time window (profile window). 
Specifically, it uses Intel PEBS~\cite{pebs} to monitor \texttt{MEM\_INST\_RETIRED.ALL\_LOADS} and \texttt{MEM\_INST\_RETIRED.ALL\_STORES}.
These events report the virtual address of the page on which the event was generated~\cite{perfmon}. 
\skd applies the Waterfall or analytic data placement model on the collected hotness profile to decide the destination tiers for the memory regions. Based on the model's outcome, \skd uses the kernel APIs described above to manage memory placement.

\sepblock
\noindent \textbf{Regions.}
In order for efficient management of the address space of an application, \skd operates at a granularity of 2\,MB regions instead of 4\,KB pages as commonly followed in other memory tiering solutions~\cite{hemem}. The hotness of 2\,MB region is an accumulated value of the hotness of each 4\,KB page in it. \skd performs data migration to and from compressed tiers at the granularity of 2\,MB regions. 


\subsection{Data placement models}
\textbf{Waterfall model:}
We implement the waterfall model in the \skd. 
The input to the model is a hotness threshold value -- $H_{th}$.
The value controls the pages that are to be evicted from DRAM to Tier 1. 
\skd maintains the tier data for all the regions and uses it to waterfall (demote to the next tier) the regions at the end of a profile window.
A region restarts its journey from DRAM when it has (or a major portion of it) been faulted back to DRAM.  

\sepblock
\textbf{Analytical Model}
We implement the analytical model in C++ using the OR-Tools from Google~\cite{ortools}. 
The input to the model is the hotness profile of the application, tier stats (e.g., compressibility ratio, cost of the media backing the compressed tier, and access latency), list of regions, and a value for the knob $(\alpha)$. The model outputs a recommendation with a destination tier for each region. 
We evaluate the model on a separate client system connected via a local network that uses socket communication to send and receive data. 

\section{Evaluation}
\label{sec:evaluation}

\begin{table}
\footnotesize
    \centering
    \caption{The set of compressed tiers used to evaluate \ntier.}
    \label{tab:fivetiersetup}
    \begin{tabular}{|l|C{1.4cm}|c|c|C{1cm}|}
    \hline
    \textbf{ID} & \textbf{Name} & \textbf{Pool manager} & \textbf{Compressor} & \textbf{Media} \\ \hline
T1 &ZB-L4-DR &zbud &      lz4       & DRAM      \\ \hline 
T2 &ZB-L4-OP &zbud &      lz4       & Optane      \\ \hline 
T3 &ZS-L4-OP &zsmalloc &  lz4       & Optane     \\ \hline 
T4 &ZS-LO-DR &zsmalloc &  lzo   & DRAM        \\ \hline 
T5 &ZS-DE-OP & zsmalloc &  deflate   & Optane \\ \hline 
    \end{tabular}%
\end{table}

\subsection{Configurations}
\label{sec:impl_tier_config}

\sepblock
\textbf{Tiers.} For evaluating \ntier, we use DRAM + 5 compressed tiers identified in Section~\ref{sec:tiers_selection} -- a total of 6 tiers.
Table~\ref{tab:fivetiersetup} shows the configuration of the 5 compressed tiers.
For evaluating the \twotier system, we use DRAM + one compressed tier, where the configuration for the compressed tier is the one employed by Google in their production data centers -- zsmalloc as the pool manager, lzo as the compression algorithm, and DRAM as the backing storage~\cite{gswap}.

\sepblock
\textbf{\skd.} We use 120 seconds as our profile window duration (as used in the state-of-the-art \twotier technique~\cite{gswap}). We observe that a time window of 120 seconds is sufficient to stabilize the hotness profile of the pages based on events generated by the hardware counters. In addition, this time window provides ample opportunity for \skd to implement the model's page placement recommendations with minimal interruptions to the applications. Each run has a warm-up window of 100 seconds. 


\sepblock
\textbf{Hotness profile.} The hotness of a region is based on the number of PEBS~\cite{pebs} samples observed during a profile interval.
For the evaluation of \twotier system and Waterfall model, we experiment with three different hotness threshold values (regions with access counts less than the threshold value are eligible for placement in a compressed tier). The threshold values are selected to cover around 15-20\% (conservative), 40-50\% (moderate), and 70-80\% (aggressive) of the application's data pages. For example, \memcached uses a threshold value of 50, 100, and 250, respectively.
Hotness statistics are gathered for pages in DRAM; hotness is not relevant for pages in compressed pools since they need to be first decompressed before accessing.
For the analytical model, the average hotness value of the region for the past 4 profiling windows
is directly fed into the model.

\sepblock
\textbf{Model configuration.}
We use the following \twotier and \ntier configurations for our evaluation.
\begin{itemize}
    \item \textbf{2T} (\twotier system): We experiment with 3 different configurations: conservative (2T-$C$), moderate (2T-$M$) and aggressive (2T-$A$) based on the hotness threshold value. 
    \item \textbf{6T-WF} (6-tier waterfall model): We evaluate with the same hotness threshold values used above for \twotier setup (6T-WF-C, 6T-WF-M, and 6T-WF-A).
    \item \textbf{6T-AM-{$\alpha$}} (6-tier analytical model): We evaluate with 3 different values of  \userinp: 0.9, 0.5, and 0.1.
\end{itemize}

\subsection{Experiment setup}
\label{sec:eval_setup}
We use a tiered memory system with Intel Xeon Gold 6252N with 2 sockets, 24 cores per socket, and 2-way HT for a total of 96 cores. It has a DRAM-based near-memory tier with 384 GB capacity and a far-memory tier with Intel’s Optane DC PMM~\cite{optane} configured in flat mode (i.e., as volatile main memory) with 1.6 TB capacity. We run Fedora 30 and use a modified Linux kernel, 5.17.

\begin{table}
    \centering
    \footnotesize
    \caption{Description of the workloads and configurations.}
    \label{tab:workloads}
        \begin{tabular}{|l|p{3.9cm}|C{1.6cm}|}
            \hline
            \textbf{Workloads}                        & \textbf{Description}                                                               & \textbf{Input}                    \\ \hline
            \memcached~\cite{memcached} & A commercial in-memory object
caching system. & 44\,GB, Value size: 4\,KB \\ \hline
            \redis~\cite{redis} & A commercial in-memory key-value store. & 41\,GB, Value size: 4\,KB \\ \hline
            \bfs~\cite{ligra}                         & {Traverse graphs generated by web crawlers. Use breadth-first search.}  & Nodes: 100\,M Size: 18\,GB\\ \hline
            \pagerank~\cite{ligra}                    & {Assign ranks to pages based on popularity (used by search engines).}   &  Nodes: 100\,M Size: 18\,GB       \\ \hline
            XSBench~\cite{xsbench} & A key computational kernel of
the Monte Carlo neutron transport algorithm & Setting: XL Size: 119\,GB \\ \hline
           
        \end{tabular}%
    
\end{table}

Table~\ref{tab:workloads} shows the real-world benchmarks and their configuration used to evaluate \ntier.
We initialize \memcached and \redis databases with $\approx$42\,GB of key-value pairs and then generate the load in a Gaussian distribution to better mimic real-life use cases~\cite{gaussian}. We use the widely used \memtier workload generator for load generation~\cite{redis,memcached_memtier}.
We use \pagerank and \bfs from the Ligra suite of graph benchmarks~\cite{ligra}. Input graphs for both graph workloads are generated using the standard rMat graph generator~\cite{rmat_gen}. 
We also use \xsbench which is a key computation kernel of the Monte Carlo neutron transort algorithm~\cite{xsbench}. We use the ``XL'' setting of the workload, which generates a memory footprint of 119\,GB.

\begin{figure*}
\centering
    \subfloat[Memcached]{
    \includegraphics[width=.25\linewidth]{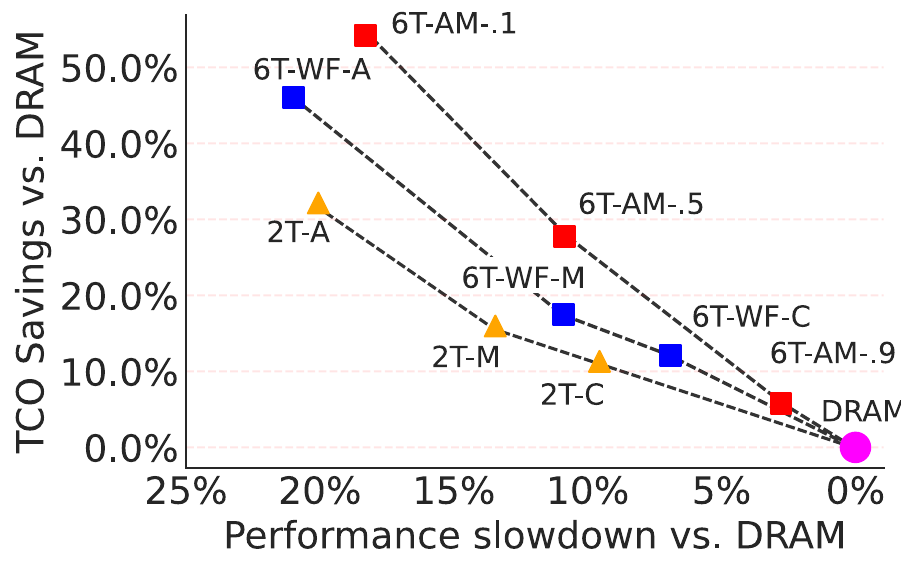}
    }
    \subfloat[Redis]{
    \includegraphics[width=.25\linewidth]{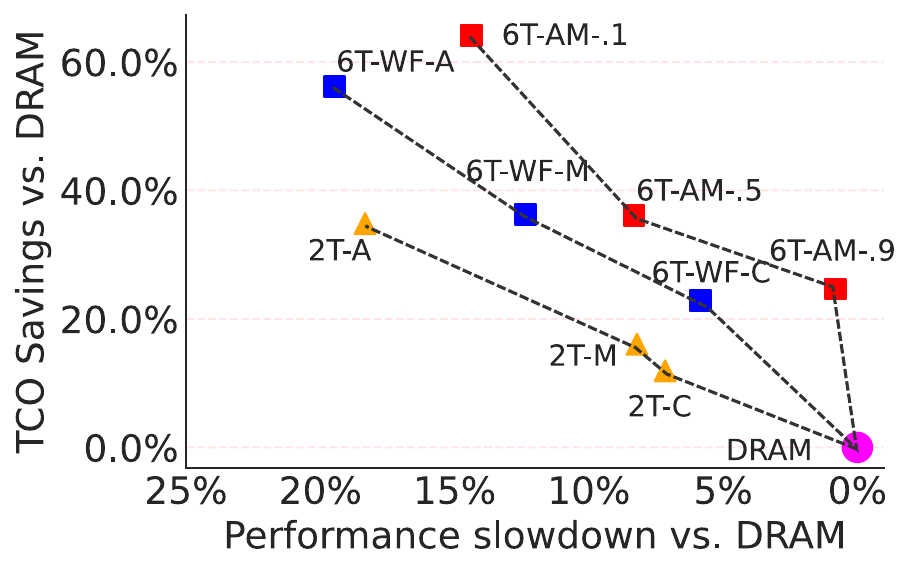}
    }
    \subfloat[PageRank]{
    \includegraphics[width=.25\linewidth]{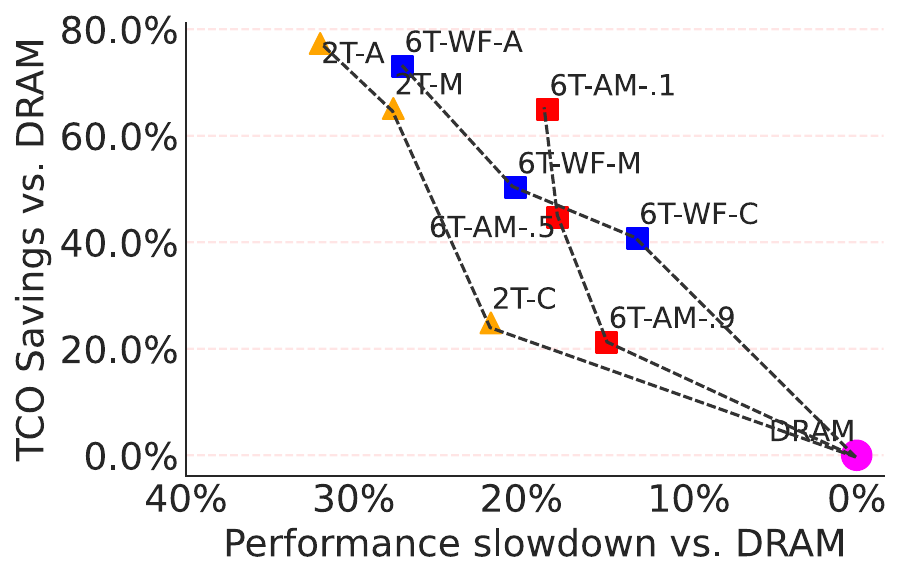}
    }
    
    \subfloat[BFS]{
    \includegraphics[width=.25\linewidth]{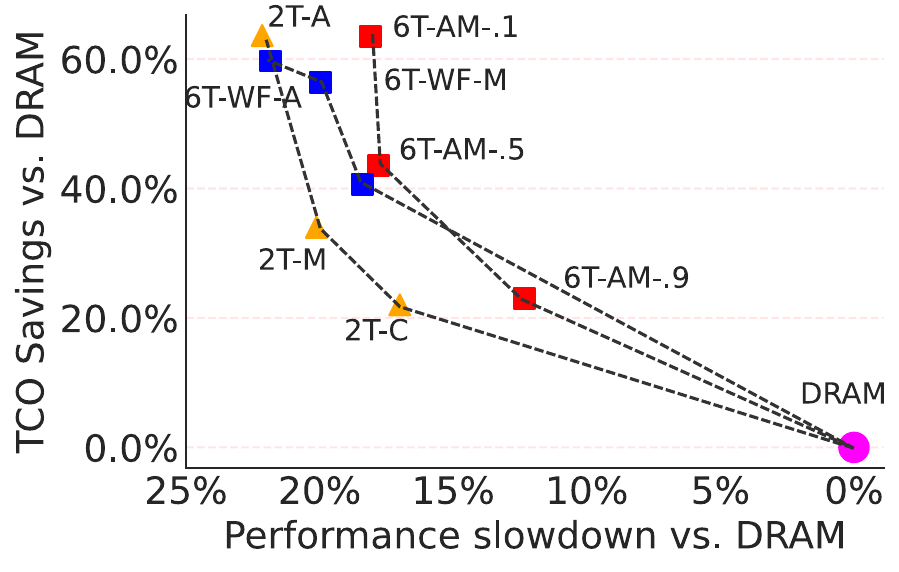}
    }
    \subfloat[XSBench]{
    \includegraphics[width=.25\linewidth]{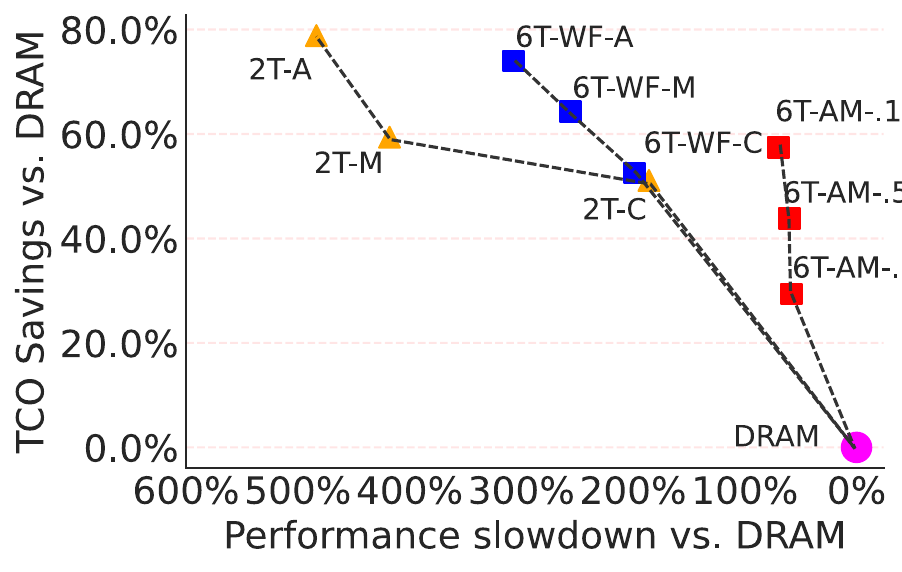}
    }
\caption{Performance slowdown and memory TCO savings w.r.t to DRAM for 2-Tier and TierScape solutions}
\label{fig:star_plots}
\end{figure*}

We execute the benchmarks and place the data as recommended by our data placement models.
For \memcached and \redis, we use the throughput and latency numbers reported by \memtier. For \pagerank and \bfs, we report the geometric mean of the time taken to execute multiple rounds. For \xsbench, we use the time reported by the benchmark.

To calculate the memory TCO we use Equation~\ref{eq:tcocost}. We capture the resident set size (RSS) to compute the cost of storing pages in DRAM and capture the size of compressed memory in all the tiers to compute the cost of storing data in the respective compressed tiers. We set the per-GB cost of Optane as 1/3 of DRAM~\cite{flexhm}.

\subsection{Results}
\label{sec:eval_performance_comparison} 
Figure~\ref{fig:star_plots} compares relative performance and memory TCO savings for 2-Tier and TierScape systems on
different workloads. Note that values on the x-axis are on a decreasing scale.

\subsubsection{TCO savings}
It can be observed from Figure~\ref{fig:star_plots} that \ntier waterfall and analytical models outperform \twotier solutions by saving more memory TCO with similar or better performance.
For example, the maximum memory TCO savings a \twotier solution can offer for \redis is 34.84\% with the 2T-A configuration. But it suffers a 18.33\% performance slowdown. \ntier's waterfall model using the same hotness threshold (6T-WF-A) achieves a TCO saving of 56.11\% (21.27 percentage points better than 2T-A) while incurring a performance loss of 19.48\% (only 1.15 additional percentage points than the 2T-A).
The analytical model configuration 6T-AM-0.1 achieves a TCO savings of 64.10\% (29.26 more percentage points than 2T-A) while incurring a performance loss of only 14.37\% (3.96 fewer percentage points than 2T-A).


\subsubsection{Performance overheads}
Similarly, it can be observed from Figure~\ref{fig:star_plots} that \ntier waterfall model and analytical model outperform \twotier solutions by performing with similar or better memory TCO savings.
For example, PageRank, in a \twotier system, 2T-C offers the least performance slowdown of 21.82\% while saving 24.86\% memory TCO. \ntier waterfall model using the same hotness threshold (6T-WF-C) offers a better trade-off than 2T-C with 13.09\% performance slowdown and 40.78\% memory TCO savings.
6T-AM-0.9 incurs performance slowdown of only 14.91\% 
and offers a TCO savings of 21.26\% 

Also, it can be observed in BFS that 6T-AM-.1 and 2T-M result in around 63\% memory TCO savings, but 6T-AM-.1 performs better by 4.05 percentage points (22.15\% vs. 18.10\%). This demonstrates that with \ntier, more warm pages can be placed in compressed tiers to achieve better memory TCO savings without hurting performance.
 

\subsection{Waterfall vs. Analytical model}
\label{sec:waterfall_vs_ilp}

\begin{figure}[!htb]
    \centering
    \subfloat[6T-WF-C \label{fig:wf_c_placement}]{
    \includegraphics[width=.31\linewidth]{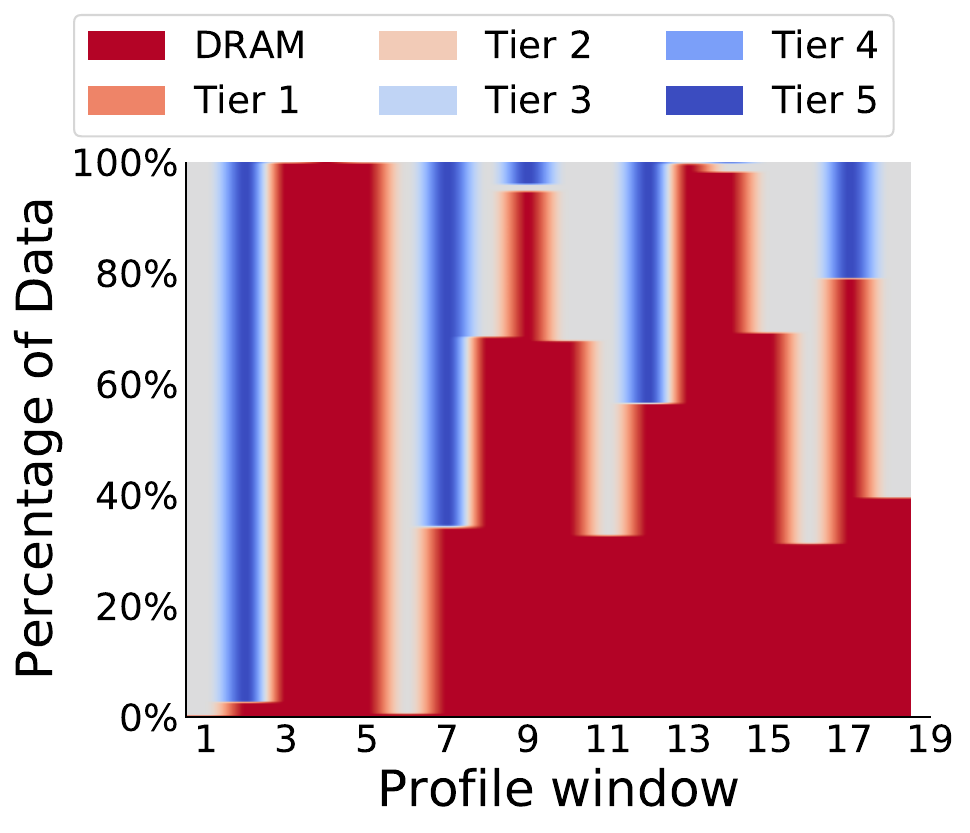}
    }
        \subfloat[6T-WF-M \label{fig:wf_a_placement}]{
    \includegraphics[width=.31\linewidth]{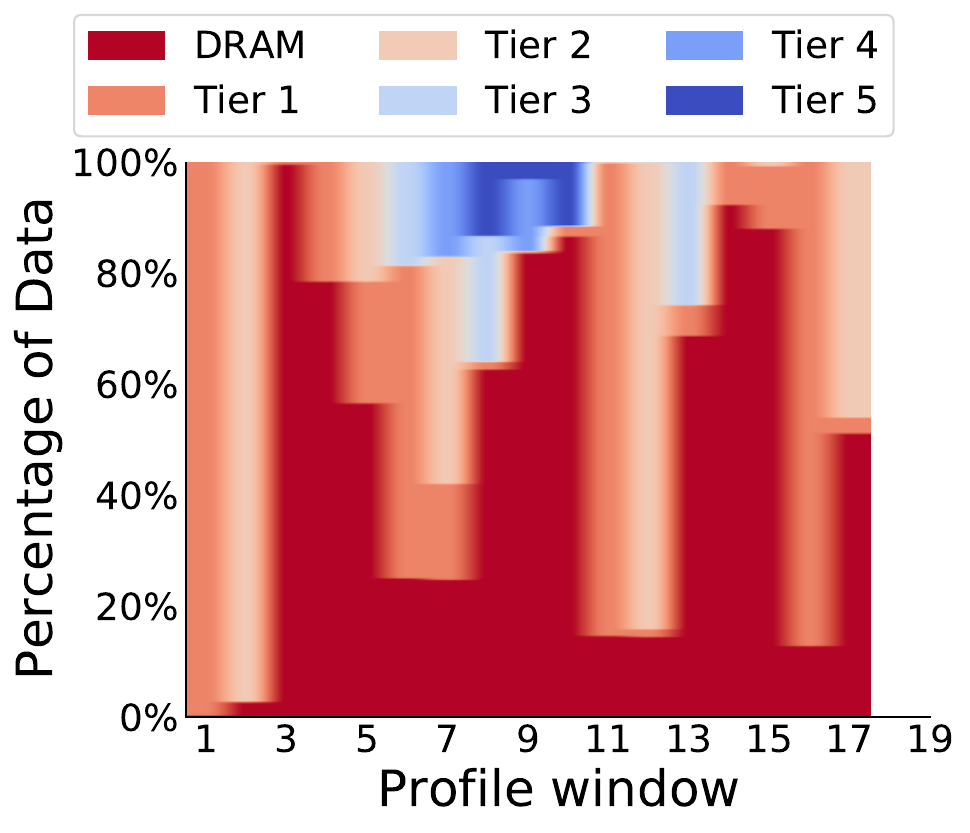}
    }
    \subfloat[6T-WF-A \label{fig:wf_ha_placement}]{
    \includegraphics[width=.31\linewidth]{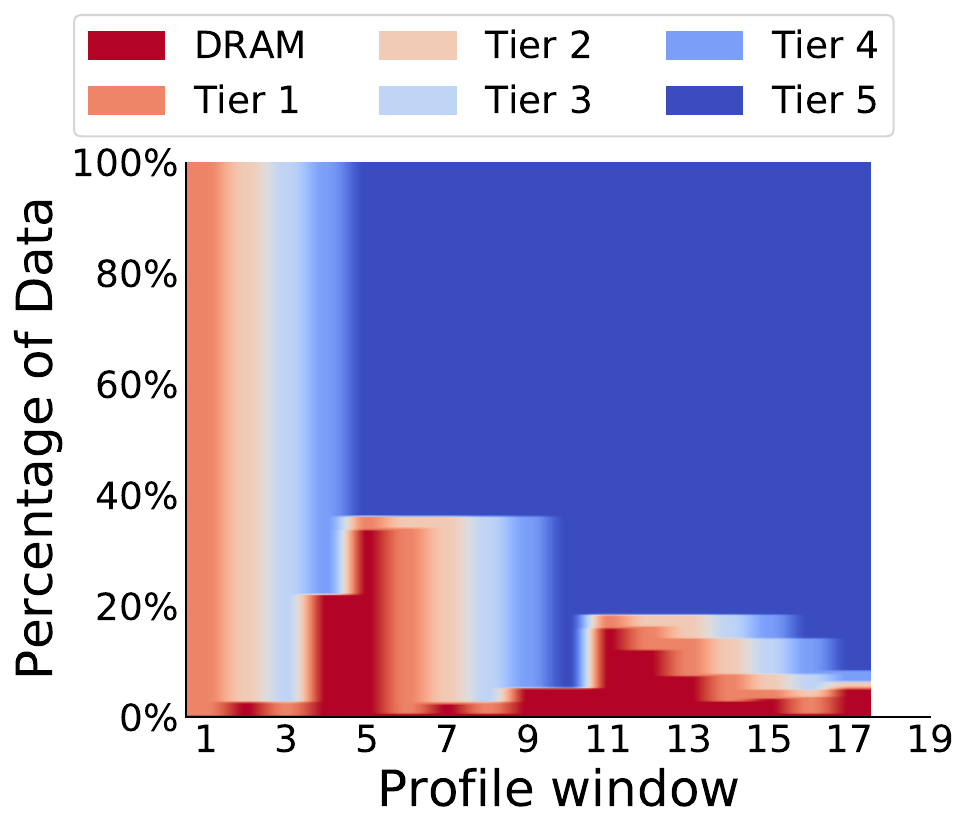}
    }

        \subfloat[6T-AM-0.9 \label{fig:am_c_placement}]{
    \includegraphics[width=.31\linewidth]{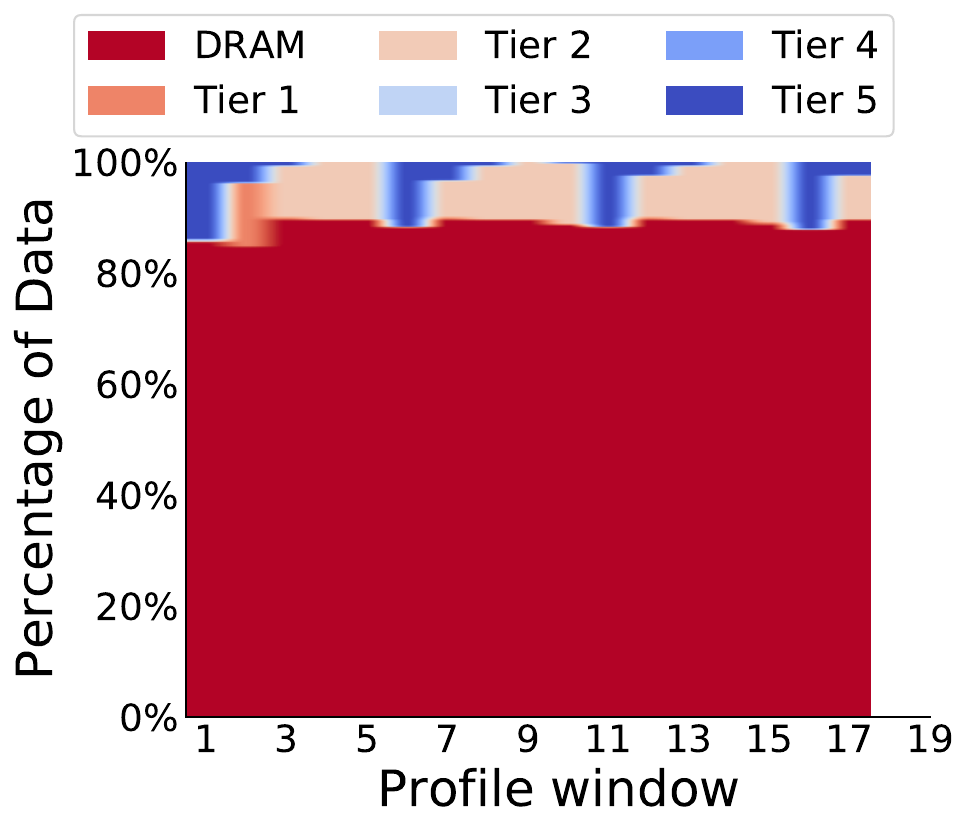}
    }
    \subfloat[6T-AM-0.5 \label{fig:am_a_placement}]{
    \includegraphics[width=.31\linewidth]{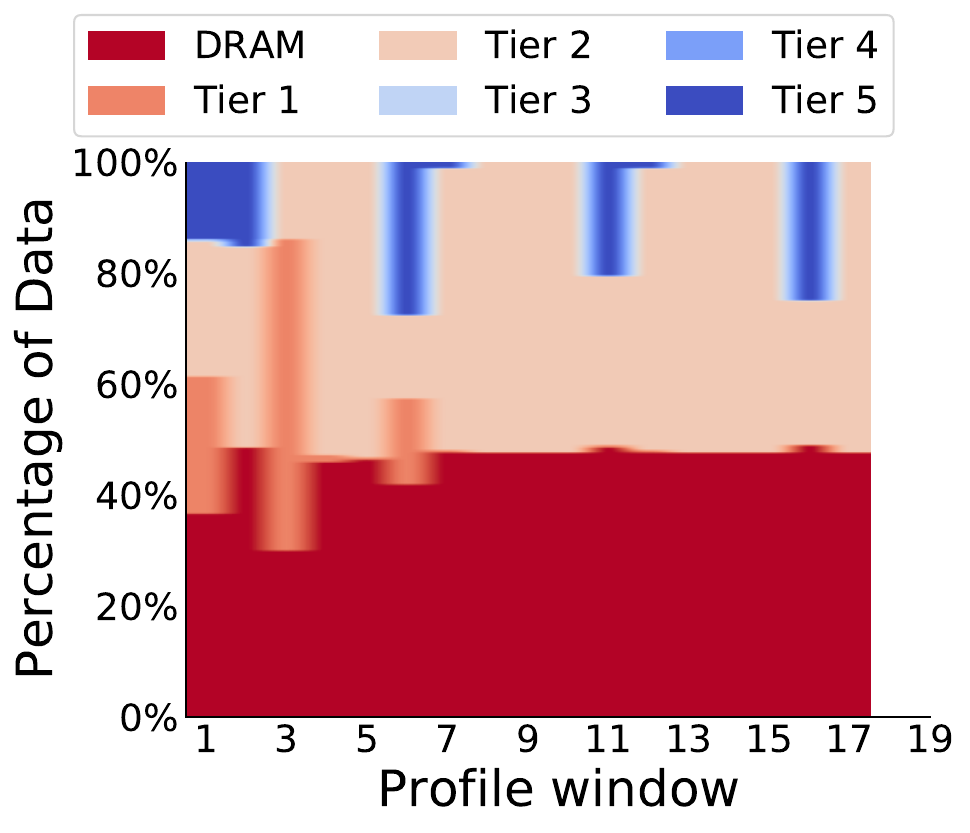}
    }
    \subfloat[6T-AM-0.1 \label{fig:am_ha_placement}]{
    \includegraphics[width=.31\linewidth]{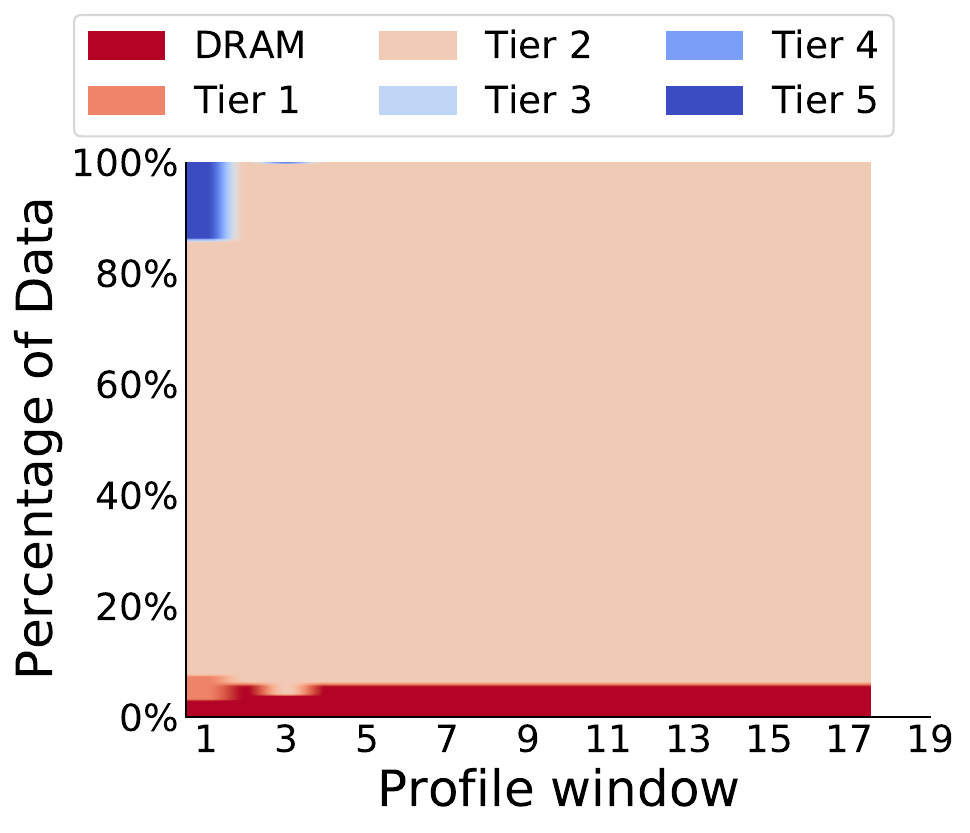}
    }
    
    \caption{Data placement recommendations for \memcached by waterfall and analytical models} 
    \label{fig:data_placement_recommendation}
\end{figure}

In this section, we deep dive and analyze the waterfall and the analytical model's data placement recommendation. Figure~\ref{fig:data_placement_recommendation} shows the model's recommendation in each profiling window for \memcached. 

6T-WF-C and 6T-AM-0.9 retain most of the data in DRAM to ensure minimal performance slowdown as per the configured hotness threshold and tunable knob values. Both models recommend placing a small amount of data in compressed tiers. 6T-AM-0.9 consistently recommends retaining more than 80\% of data in DRAM for all profiling windows.

6T-WF-M and 6T-AM-0.5 recommend placing more pages in compressed tiers. In the waterfall model, more pages are ``waterfalled'' to tiers with better compression ratio, thus increasing the utilization of all the compressed tiers. In the analytical model, the percentage of pages that are retained in DRAM is 50\% as the model converges towards placements with greater memory TCO savings based on the input to the model.
The analytical model periodically recommends scattering many pages to the best TCO-saving tier, i.e., Tier 5.

6T-WF-A, with aggressive memory TCO savings, retains only around 10\% of data in DRAM. We see a significant jump in the utilization of the last tier. 
This is because more data is swapped out from DRAM and are ``waterfalled'' between the tiers, eventually reaching the last tier.
Note that the figure shows the model's recommendation of the data placement based on the page hotness in the previous profile window. 
However, we observe that \memcached faults upon some of these pages, which are immediately moved back to DRAM. 

6T-AM-0.1 recommends placing less than 5\% of data in DRAM and the rest in compressed tiers. It recommends placing a majority of the data in Tier 2 instead of Tier 5. Tier 2 in our setting is zbud with lz4 backed by Optane. We observe that the last tier, using deflate, maintained an average compressibility ratio of 2 for \memcached. Whereas Tier 2 using lz4 achieved an average compressibility ratio of 1.35. 
Based on the overall cost of storing data and the underlying TCO, the model decided that placing most of the data on Tier 2 satisfies the TCO constraints. It can be noted that the model still recommends placing a small amount of data in the rest of the tiers.

\begin{figure*}
    \centering
    \subfloat{
    \includegraphics[width=.5\linewidth]{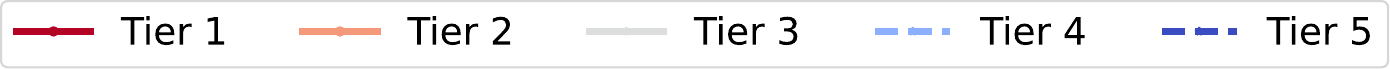}
    }
   
    \subfloat[Pages in 6T-WF-M]{
    \includegraphics[width=.2\linewidth]{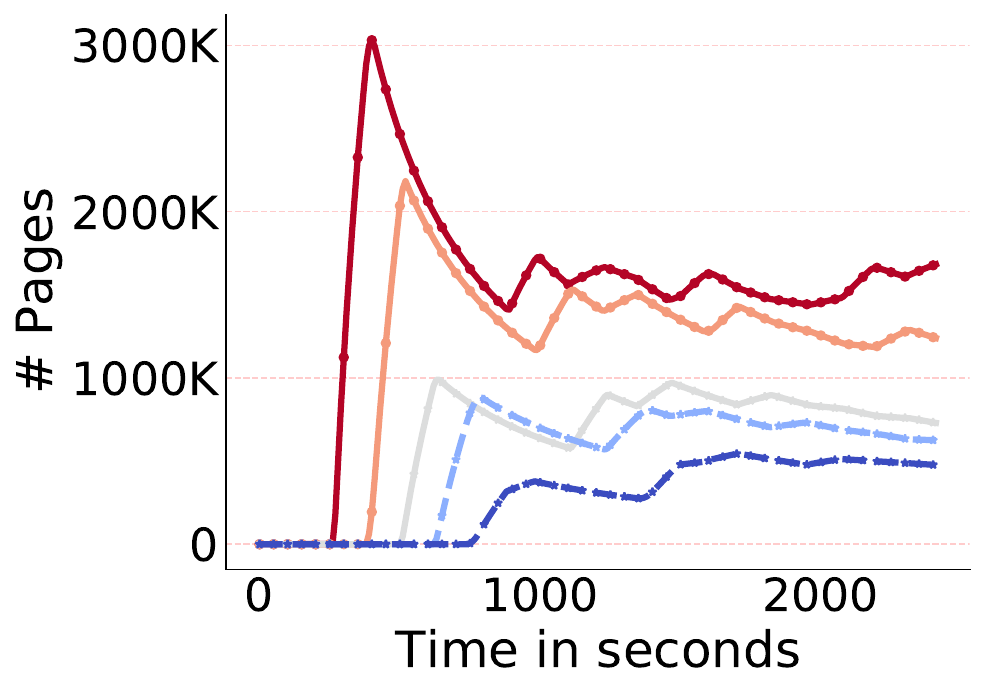}
    }
    \subfloat[Pages in 6T-WF-A]{
    \includegraphics[width=.2\linewidth]{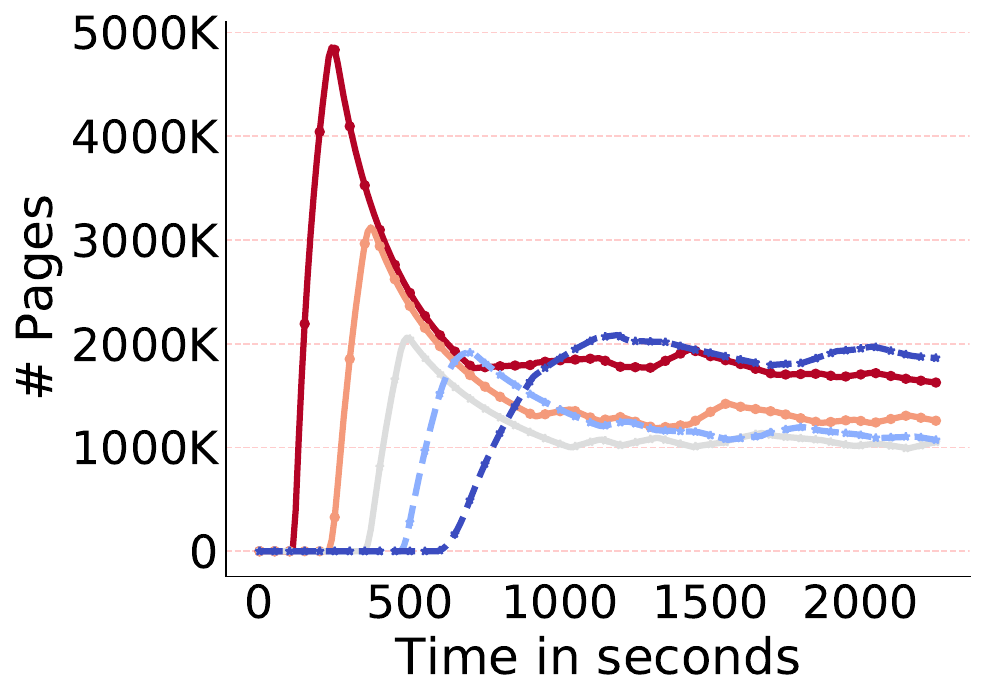}
    }
    \subfloat[Pages in 6T-AM-.5]{
    \includegraphics[width=.2\linewidth]{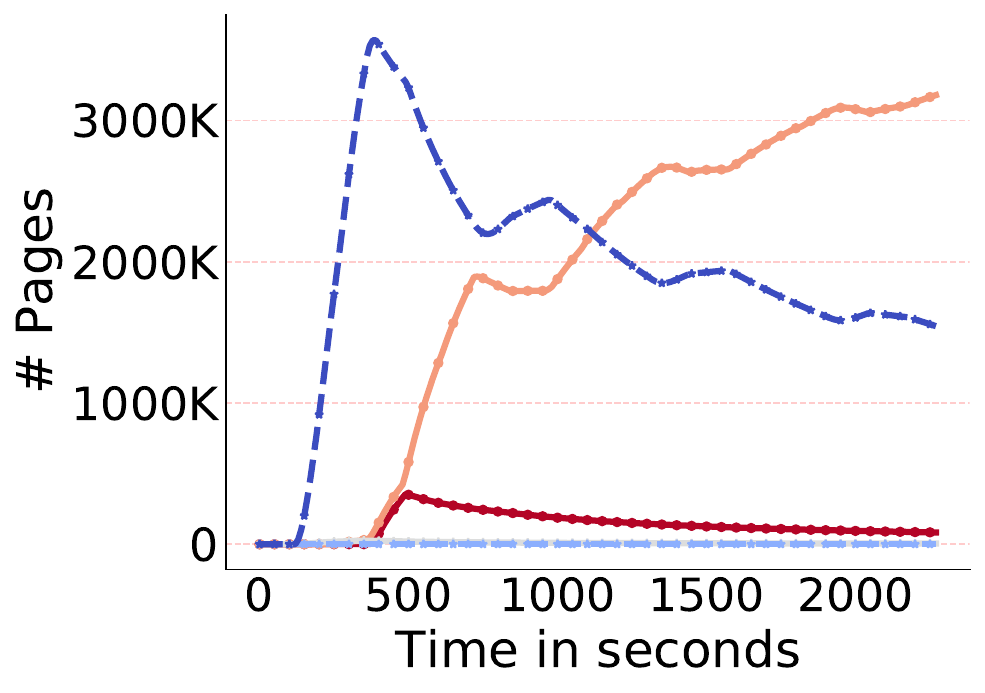}
    }
    \subfloat[Pages in 6T-AM-.1]{
    \includegraphics[width=.2\linewidth]{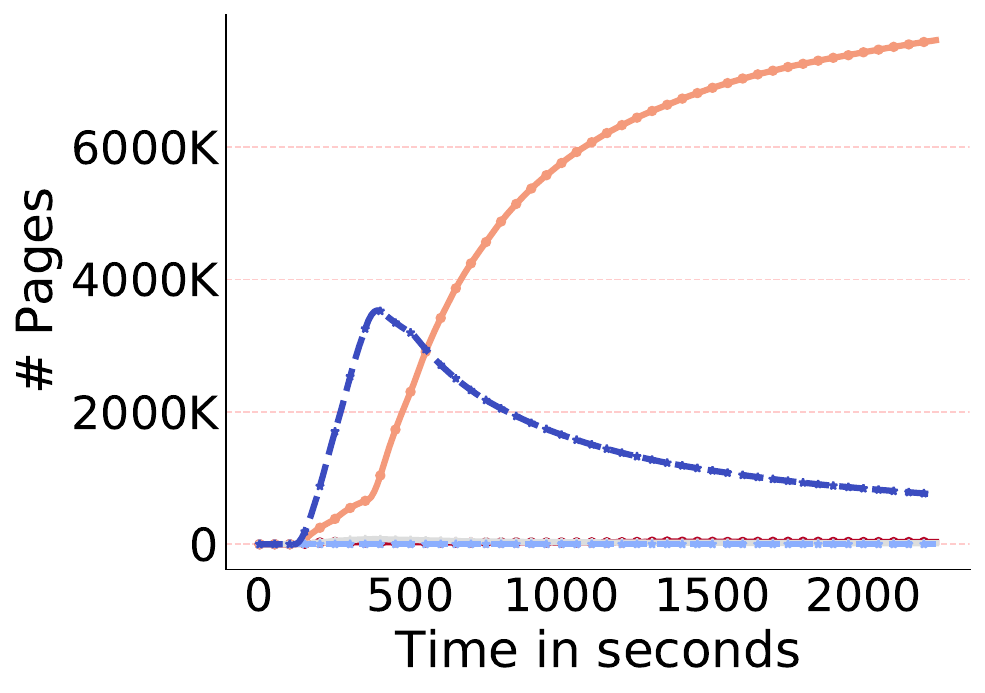}
    }
    
   \subfloat[Faults in 6T-WF-M]{
    \includegraphics[width=.2\linewidth]{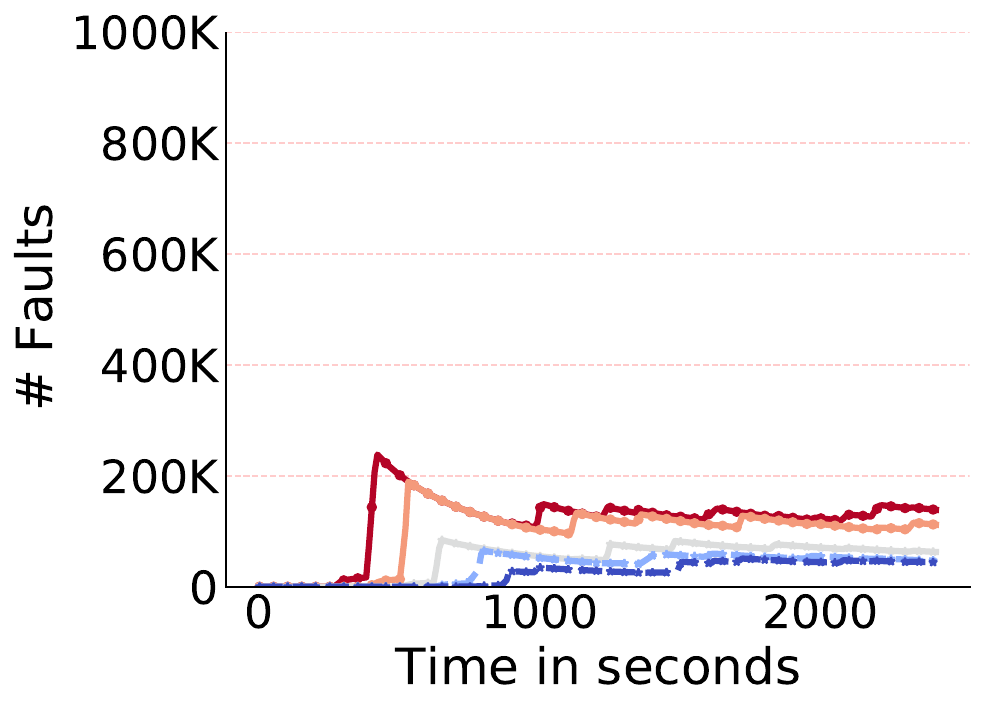}
    }
    \subfloat[Faults in 6T-WF-A]{
    \includegraphics[width=.2\linewidth]{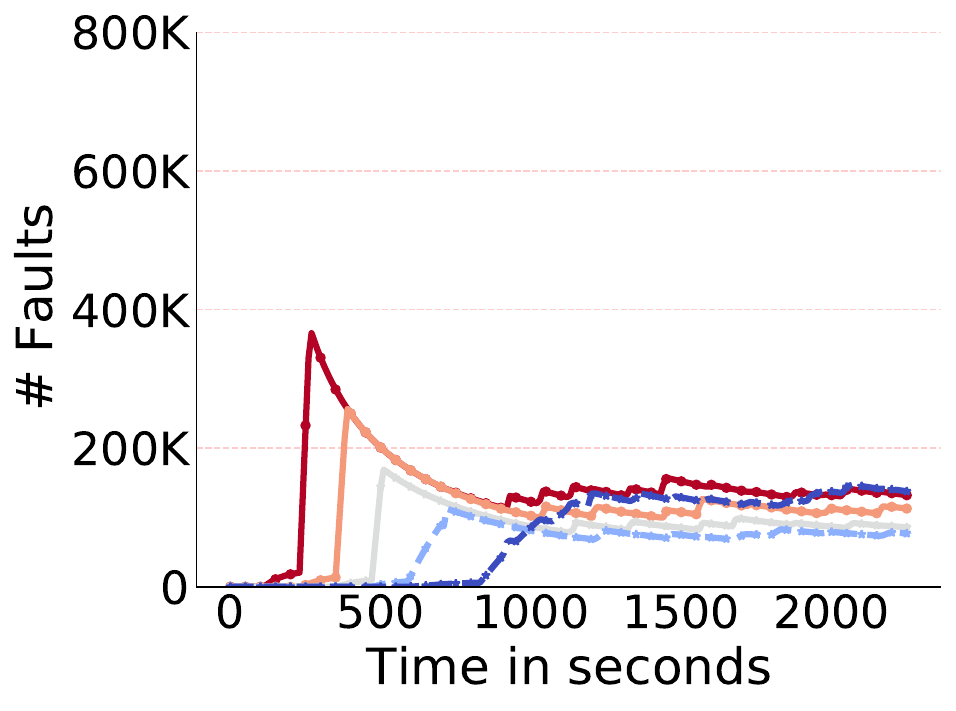}
    }
    \subfloat[Faults in 6T-AM-.5]{
    \includegraphics[width=.2\linewidth]{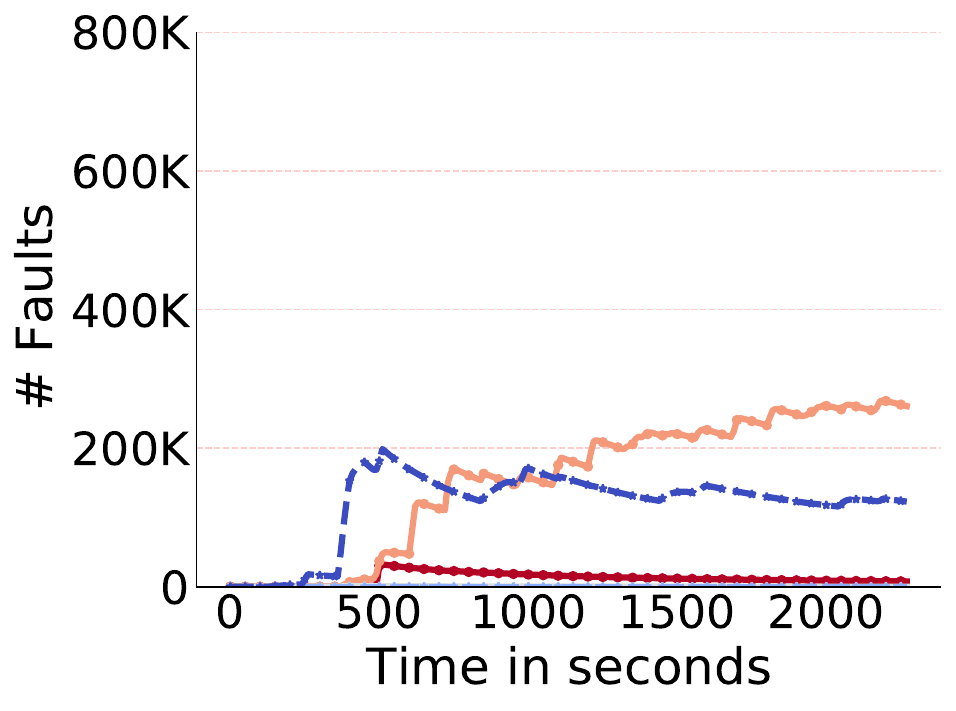}
    }
     \subfloat[Faults in 6T-AM-.1]{
    \includegraphics[width=.2\linewidth]{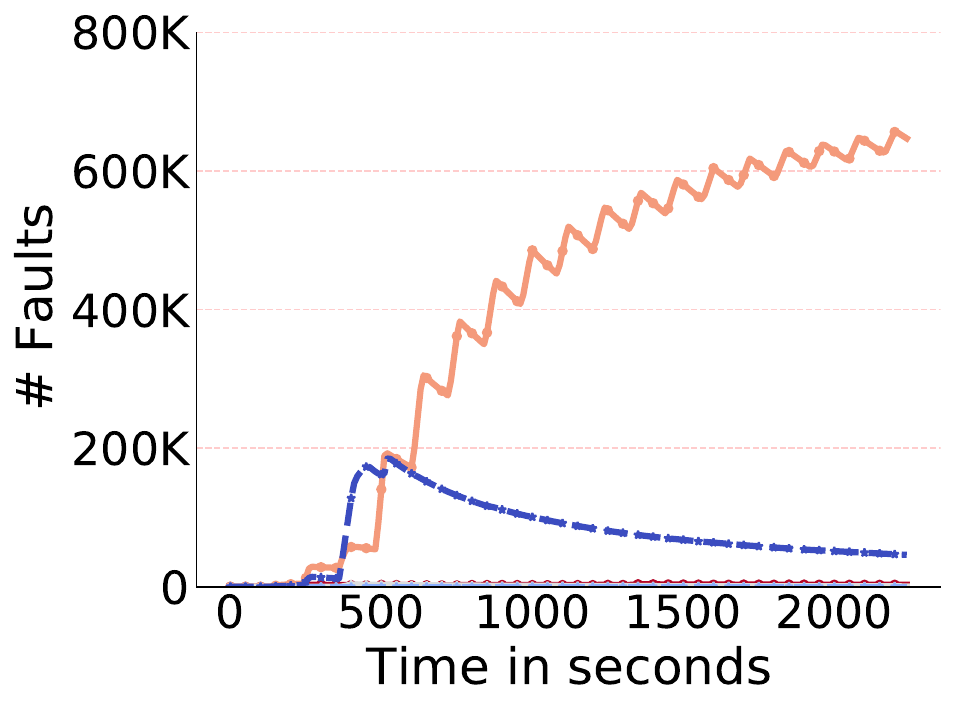}
    }
    
    \caption{Pages placement across tiers by \skd as per the recommendation by the waterfall and analytical models and the actual page faults observed for \redis benchmark. Please note the difference in y-axis scale across plots.
    }
    \label{fig:redis_pages_faults}
\end{figure*}

\begin{figure}
    \centering
    \subfloat[Waterfall model]{
    \includegraphics[width=.4\linewidth]{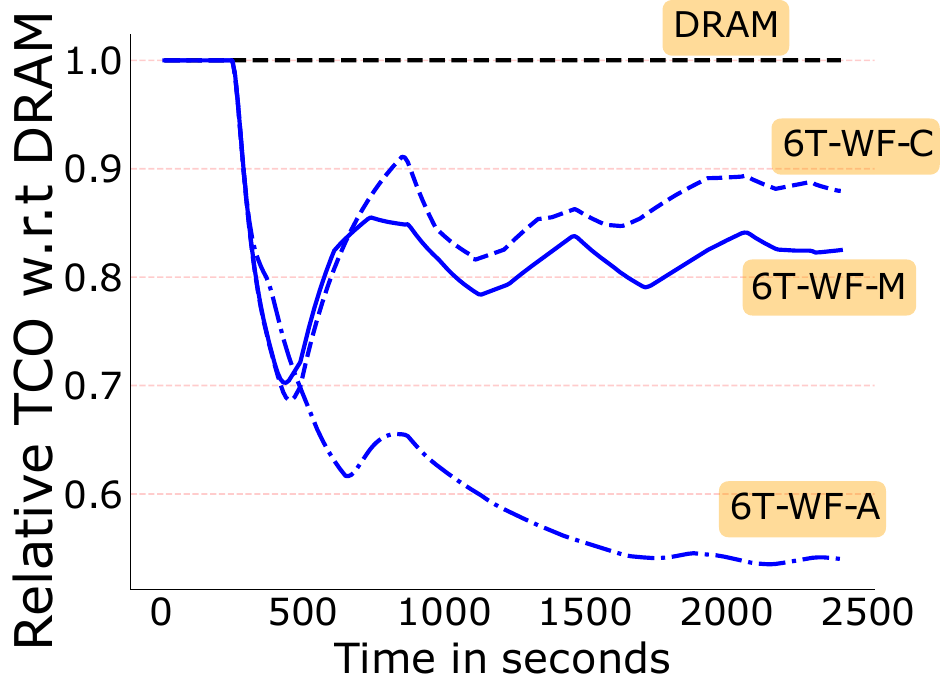}
    }
    \subfloat[AM model]{
    \includegraphics[width=.4\linewidth]{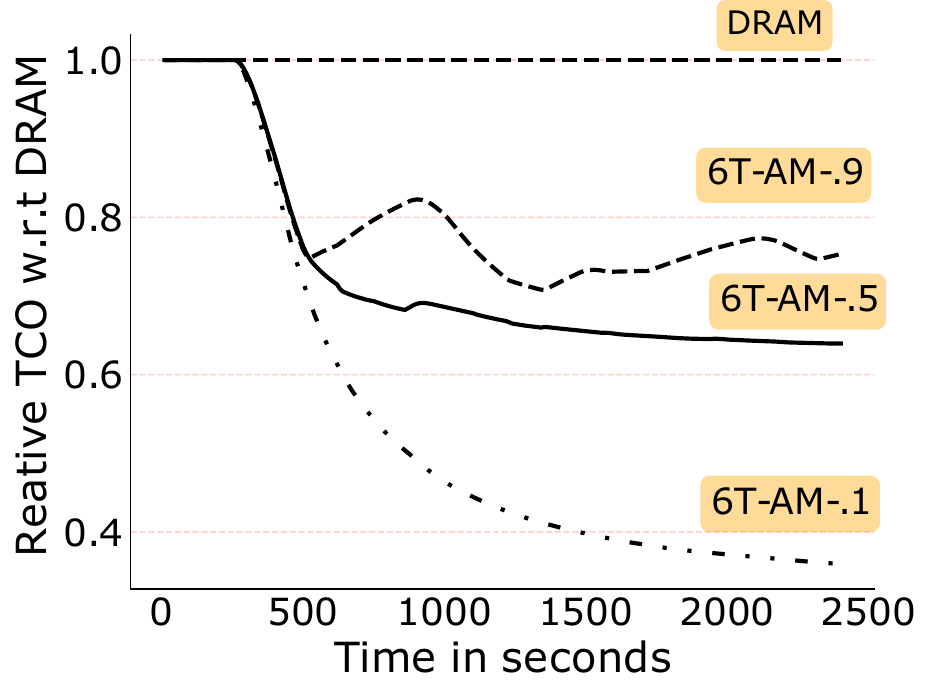}
    }
    \caption{Memory TCO savings for \redis} 
    \label{fig:tco_trend}
\end{figure}

\subsection{In-depth Analysis}
\label{sec:in_depth_analysis}

In the previous section, we looked into the model recommendation, while in this section, we analyze the ground reality, i.e., the number of pages actually placed in multiple tiers by \skd as per model recommendation and the on-demand page faults incurred by the applications.
Figure~\ref{fig:redis_pages_faults} shows the data placement for \redis
for 6T-WF-M, 6T-WF-A, 6T-AM-.5, and 6T-AM-.1. We omit other benchmarks and configurations as they show a similar trend.

For the waterfall model, it can be observed from Figure~\ref{fig:redis_pages_faults} that after initial warm-up time, pages are first ``waterfalled'' to Tier 1, and then they are gradually aged to better memory TCO saving tiers.
The difference in placement recommendation between 6T-WF-M and 6T-WF-A is clearly visible, where 6T-WF-M (moderate configuration) starts by placing around 3000K pages in Tier 1, while 6T-WF-A (aggressive configuration) starts by placing around 5000K pages in Tier 1.
In 6T-WF-A all the tiers are almost equally utilized after a few profile windows (1000 seconds onward). While in 6T-WF-M the fast tiers (Tier 1 and Tier 2) have more pages than the slow tiers. This is as per the ``moderate'' configuration, which balances performance impact and memory TCO savings. 

In the analytical model, 6T-AM-.5 starts with placing pages in Tier 5 (Figure~\ref{fig:redis_pages_faults}). It should be noted that the input to the analytical model is the average hotness of the regions for the last four profiling windows. Hence, as the model gets more information on application access pattern trends over subsequent profile windows, it starts preferring Tier 2 instead of Tier 5. A similar trend of initially placing pages in Tier 5 is also observed for 6T-AM-.1. However, 6T-AM-.1 eventually places a significantly higher number of pages in Tier 2 (7,000K pages compared to 3,000\,K pages in 6T-AM-.5), resulting in better memory TCO savings.

Further, it can be observed in the bottom plots of Figure~\ref{fig:redis_pages_faults} that only a small fraction of the pages (around 10\%) placed in compressed tiers are accessed by the application, resulting in a page fault. This clearly indicates the efficiency of waterfall and analytical models to correctly recommend the placement of the pages in the appropriate compressed tiers.
In addition, the memory TCO trend in Figure~\ref{fig:tco_trend} corroborate 
with the page placements in  
Figure~\ref{fig:redis_pages_faults}
that also reflects the placement decisions made by models with different configurations.




\subsection{Impact on the tail latencies}
\label{sec:eval_tail_latency}
One of the key requirements in a data center is to maintain an SLA guarantee on the tail latencies of an application. A hosted application should not suffer exorbitantly high tail latencies in the pursuit of aggressively reducing memory TCO. 
In N-tier systems, the total number of faults in the slowest tier 
and the decompression (or access) latency of the slowest tier can impact the tail latency of the application.

\begin{figure}
    \centering
  \includegraphics[width=.8\linewidth]{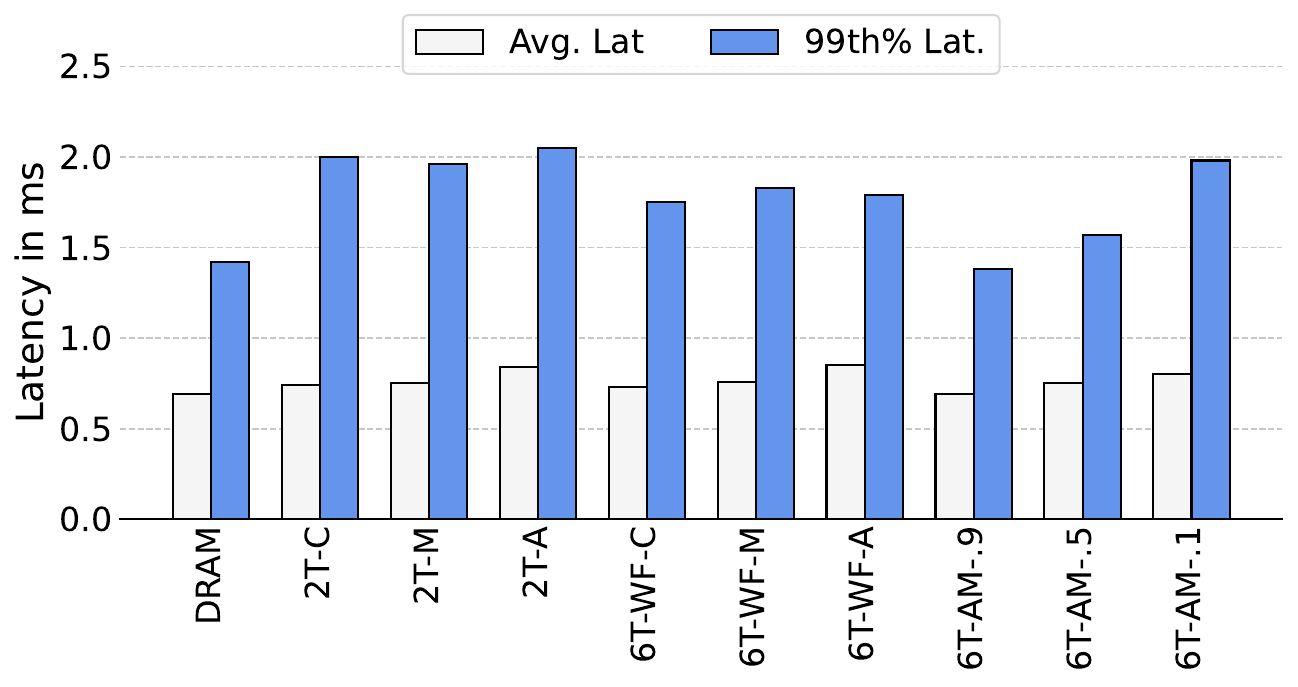}
    \caption{Latency data for \memcached}
    \label{fig:memcached_latency}
\end{figure}

Figure~\ref{fig:memcached_latency} shows
the average and 99th percentile latency for Memcached. 
It can be observed
that the latency values, both average and 99th percentile, increases for both 2-Tier and N-tier (6T-WF and 6T-AM) as we aggressively place more pages in compressed tiers. The average latency values for 2T, 6T-WF and 6T-AM are comparable for similar aggressive settings (e.g., 2T-C vs. 6T-WF-C).

It can be observed that 6T-WF and 6T-AM outperform all 2T configurations for 99th percentile latency. For waterfall model as pages are gradually aged into slowest tier, only pages that are actually cold end up in slowest tier and hence performs better than all 2T configurations. Analytical model carefully scatters the pages across tiers based on the hotness values of the pages in the past four profile windows. Hence the 99th percentile latency for 6T-AM-.9 and  6T-AM-.5 is better than 2T and 6T-WF. However, for 6T-AM-.1 with aggressive memory TCO savings settings, 99th percentile latency is higher than 6T-WF-A, but is still better than 2T-A.


\subsection{\skd Tax}
\label{sec:eval_tax}

In this section, we analyze the tax in terms of CPU utilization incurred by \skd.
The tax includes telemetry data collection and post-processing along with page migration tax (including decompressing a page from one tier and compressing and placing it in another tier). Pages decompressed due to on-demand faults are not accounted for in \skd as they are accounted for in the benchmark execution time. 

\begin{figure}
    \centering
    \includegraphics[width=\linewidth]{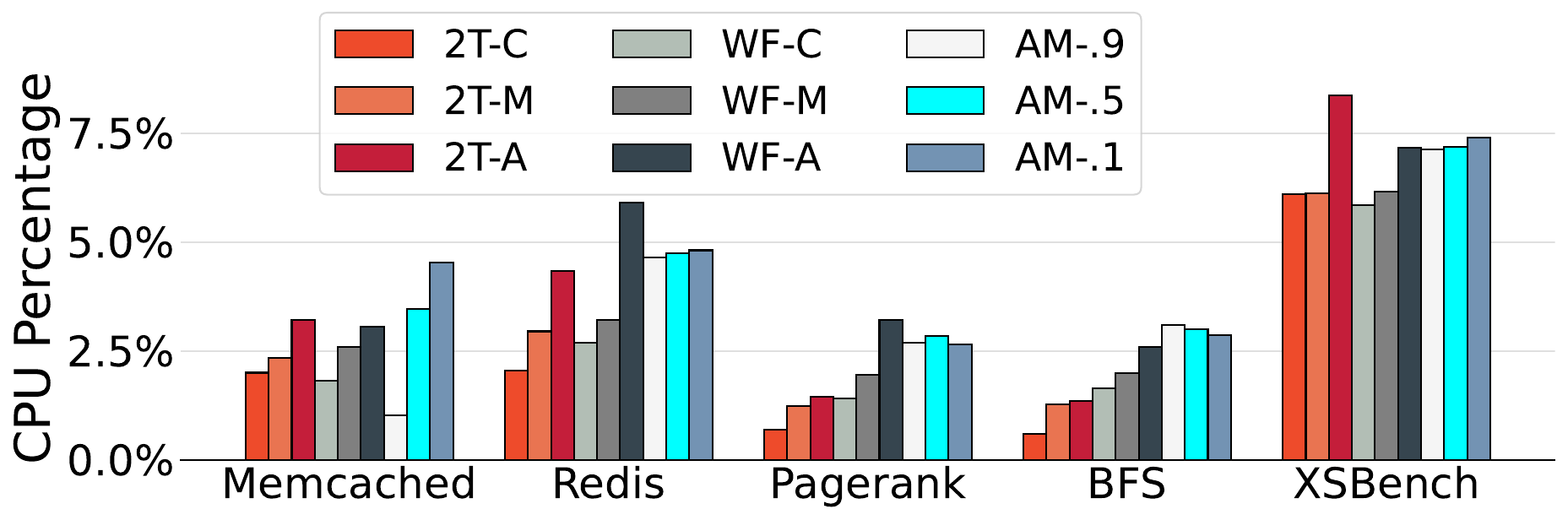}
    \caption{CPU utilization by \skd which includes telemetry collection, post-processing the telemetry data, page migration, and (de)compression overheads.}
    \label{fig:tco_tax}
\end{figure}

It can be observed from Figure~\ref{fig:tco_tax} that \skd incurs around 1.2\% to 7\% CPU utilization for different benchmarks. For most of the benchmarks, CPU utilization for \skd is slightly higher for waterfall and analytical models, as additional CPU cycles are burned during every profiling window to redistribute the pages across multiple tiers as per the recommendation by the model.

Also, the tax for the analytical model does not include the CPU overheads to evaluate the model as it is offloaded to a client system. We also measure the overhead on the client system, which contributes to less than 0.4 percentage point increase in CPU utilization for analytical models.

\section{Related Work}
\label{sec:related_work}

Several tiered memory systems have been proposed in recent years~\cite{gswap, tmo, tpp, tmts, pond, thermostat, memtis, nimble, hemem, AutoTiering, heterovisor, radiant, cxl-expansion-databases, cxl-dl, cxl-pooling}, along with data placement and migration policies to optimize performance and memory TCO. Most of the prior work are based on two tier system, where the first tier consists of low latency and costly DRAM memory while the second tier consists of high latency but cheaper memory tiers backed by NVMMs~\cite{optane} or CXL-attached memory~\cite{cxl-website, cxl-samsung, cxl-micron}. Recently, memory tiering using a compressed memory tier has been explored by a hyper-scale data center provider~\cite{gswap}.

Hardware-based memory tiering with NVMMs~\cite{nimble, thermostat, tmts, tmo, AutoTiering} or CXL-attached memory~\cite{pond, tpp, cxl-expansion-databases, cxl-dl, cxl-pooling} lack the flexibility in the software to define memory tiers with distinct access latency, as access latency is determined by the underlying storage media. While 
\ntier proposes a way to define  multiple copressed memory tiers in the software.

Prior proposals employ different telemetry techniques to identify hot and cold data~\cite{pebs, mglru, idle-page-tracking, kstaled, damon}. HeMem~\cite{hemem}  accumulates access information from PEBS~\cite{pebs} into larger regions to reduce the overheads of tracking pages at 4\,KB granularity. \skd also uses PEBS and operates at larger regions for hot, warm, and cold data tracking.


Page placement policies employed in prior works are fine tuned for page placement in a two tier systems~\cite{tmts, tmo, hemem, tpp, pond, gswap} and cannot be directly applied to N-tier systems as they do not exploit the distinct access latency and capacity savings across tiers.
\ntier proposes Waterfall and analytical models for efficient page placement across tiers.

Prior works also employ optimizations to decide the time and rate at which the pages are migrated across tiers~\cite{gswap, hemem, nimble}. 
For example, Nimble~\cite{nimble} proposes multi-threaded and concurrent migrations of pages across tiers. \skd also employs multi-threaded and concurrent migrations of pages. 



The recent work from Google~\cite{gswap} proposes a software-defined two tiered system with DRAM and a single compressed tier to improve memory TCO savings for cold data. 
The \texttt{ACCESSED} bit in the page table is periodically scanned to identify and migrate 
cold pages to a compressed memory tier backed by DRAM. An AI/ML based prefetching technique is also employed to proactively move pages from compressed second tier memory to DRAM. 
\ntier differs from Google's approach as it defines and manages multiple software-defined compressed memory tiers.

\section{Conclusion}
\label{sec:conclusion}
We conclude with comprehensive experimental evidence that defining multiple compressed memory tiers in the software and exploiting data placement across tiers is an optimistic way forward to tame the high memory TCO in modern data centers.

\bibliographystyle{plain}
\bibliography{refs}

\end{document}